\documentclass[10pt,newlfont,superscriptaddress,twocolumn,aps,pra,letter]{revtex4}
\usepackage[pdftex]{graphicx}
\usepackage[sort&compress]{natbib}
\usepackage{exscale}
\def\hs{\hspace{0.5cm}}

\def\ft{\hspace{0.1cm}}

\def\ea{{\it et al.}}
\def\he4{${}^4$He}

\def\Ip{$I_{overlap}$}
\def\Ibol{$\langle I_{BOSON-OL} \rangle$}
\pdfoutput=1
\begin{document}

\title{Tunneling of a few strongly repulsive hard-sphere bosons in an optical lattice with 
tight external harmonic confinement: A quantum Monte Carlo investigation in continuous space}
\author{Asaad R. Sakhel}
\affiliation{Al-Balqa Applied University, Faculty of Engineering Technology, Basic Sciences Department,
Amman 11134, JORDAN}
\author{Jonathan L. Dubois}
\affiliation{Lawrence Livermore National Lab, 7000 East Ave, L-415, Livermore CA 94550, USA}
\author{Roger R. Sakhel}
\affiliation{Department of Basic Sciences, Al-Isra Private University, Amman 11622, JORDAN}
\date{\today}

\begin{abstract}
\hs The effect of strongly repulsive interactions on the tunneling 
amplitude of hard-sphere (HS) bosons confined in a simple cubic (sc) optical 
lattice plus tight external harmonic confinement in continuous space is 
investigated. The quantum variational Monte Carlo (VMC) and the variational path 
integral Monte Carlo (VPI) techniques are used at zero temperature. The effects of 
the lattice spacing $\pi/k$ on the tunneling amplitude is also considered.
The occupancies of the lattice sites as a function of the repulsion between the 
bosons are further revealed. Our chief result is, that for a 
small number of bosons (N=8) the overlap of the wave functions in neighboring wells 
does not change with an increase of the repulsive interactions and changes only 
minimally for a larger number of particles ($N=40$). The tunneling amplitude rises 
with a reduction in the lattice spacing. 
In addition, the occupancy of the center of the trap decreases in favor of a rise 
in the occupancy of the lattice sites at the edges of the trap with increasing 
HS repulsion. Further, it was found 
that the energy per particle at certain optical depths is insensitive to the number 
of particles and variations in the HS diameter of the bosons. In order to support 
our results, we compare the VMC results with corresponding VPI results.
\end{abstract}

\maketitle


\section{Introduction}

\hs The tunneling of bosons in optical lattices has drawn considerable 
interest in the last few years 
\cite{Greiner:02,Schmidt:06,Huhtamaeki:07,Luehmann:08,Foelling:07,Zoellner:08,Li:06,Barankov:03,Clark:06,Blackie:07,Lin:08} 
due to the fact the investigation of the tunneling amplitude in optical lattices 
provides insight into the physics of lattice bosons. This is due to its connection 
to superfluidity \cite{Shams:09} and its analogy to Josephson tunneling
in quantum devices \cite{Saito:02}. Particularly the role of interparticle
interactions in determining the tunneling amplitude has only been given
few investigations.

\hs In simple elementary terms, the quantum tunneling of a particle through 
a potential barrier, such as in an optical lattice, can occur when its total 
energy $E$ is less than the height of the barrier $V_0$. The 
particle is described by a wave packet which can penetrate the barrier with 
a finite probability. As a result, an overlap between two wave functions at 
both sides of the barrier provides a measure for the tunneling amplitude of 
the particle. Potential barriers of this sort have been realized in quantum 
devices \cite{Saito:02} that make use of Josephson tunneling \cite{Barankov:03} and in the 
recently achieved optical trapping of bosons \cite{Yin:06}. When there is more
than a single atom in the potential well, the mechanism of quantum tunneling
is very much determined by the strength of the interatomic interactions. In
the strongly interacting regime, correlated hopping occurs 
\cite{Foelling:07,Liang:09} where atoms tunnel in pairs and competes with
single-particle tunneling.

\hs Further, one of the most important properties of single-particle 
tunneling in optical lattices is that it is a signature of a superfluid (SF) 
state \cite{Shams:09}, and if absent, of a Mott-insulator (MI) state 
\cite{Gerbier:05,Li:06,Luehmann:08}. Pair superfluidity has also been
recently discussed \cite{Liang:09}. In an MI state, single-particle
tunneling and phase coherence are absent. Consequently, the particles 
are unable to superflow, but still able to hopp from one well to the
other. This is via the correlated hopping mechanism signalled by an overlap of 
the wave functions in neighboring wells in the strongly interacting regime. 
In fact it was F\"olling \ea\ \cite{Foelling:07} who showed experimentally 
that strong interactions suppress single-particle tunneling such that 
second-order correlated tunneling is then the dominant dynamical effect. 
An SF state is quite the opposite, where considerable single-particle 
tunneling and phase coherence are observed. 

\hs In this paper, we chiefly investigate the effects of strongly 
repulsive interactions on the correlated tunneling amplitude of bosons 
in an inhomogeneous system confined by a tight combined harmonic optical 
lattice. Here, the external harmonic trap introduces an inhomogeneiety in 
the atomic density distributions. The tunneling amplitude is measured by 
the overlap integral of two wave functions in neighboring wells obtained 
from the integrated optical densities. A key point in our investigation is 
to also explore the possibility of a SF to MI transition in the presence 
of this inhomogeneiety since it has been shown that external confinement 
added to the optical lattice suppresses the MI state \cite{Gygi:06,Mitra:08}. 
We partly check this case for few-boson systems. We use quantum variational 
Monte Carlo (VMC) and variational path integral (VPI) Monte Carlo techniques in 
continuous space at zero Kelvin. To the best of our knowledge, previous work did 
not consider the effects of inhomogeneiety on the tunneling amplitude of 
strongly repulsive lattice bosons, particularly by using Monte Carlo 
techniques in continuous space.

\hs To this end, the tunneling amplitude of bosons in optical lattices has been 
investigated chiefly as a function of the optical depth (i.e., barrier height $V_0$) 
\cite{Jaksch:98,Shams:09,Foelling:07} and number of particles \cite{Li:06}. 
An investigation most relevant to our work is that of Shams and Glyde 
\cite{Shams:09}. They evaluated the BEC density and superfluid fraction of HS 
bosons confined in an external periodic potential using PIMC in order to shed 
further light on the connection between BEC and superfluidity. In part, they 
investigated the hopping parameter as a function of $V_0$, and showed that if 
$V_0$ is increased sufficiently, the condensate is localized into islands inside 
the potential wells suppressing superflow substantially. Further, they found that 
their external potential suppresses the superfluid fraction at all temperatures.
 
\hs In our investigation a key point is that, in contrast to Shams and Glyde, 
$V_0$ is kept fixed while the repulsive interactions between the bosons 
are varied. Further, the effects of these interactions on atom correlations, 
optical density, occupancy of lattice sites, onsite interaction energies, and 
the total energies are explored. In addition, the effects due to lattice spacing 
and number of particles are further revealed. 

\hs As outlined in the method section, the bosons are represented by hard 
spheres (HS) of diameter $a_c$ whose repulsive interactions can be modified by 
changing $a_c$, thereby mimicking the Feshbach resonance technique 
\cite{Timmermans:99}. 
Our chief result is that, for a small number of bosons, the overlap of wave 
functions in neighboring potential wells does not change with increasing HS 
repulsion and changes only minimally for $N=40$. The localized wave functions 
in the potential wells do not broaden with an increase in $a_c$, in contrast to 
the case of HS bosons in pure harmonic traps \cite{Dubois:01,Sakhel:02}. In 
the latter, the width of the spacial many-body wave function extends to several 
trap lengths, whereas in our case, the wavefunction in each well extends only 
slightly beyond two trap lengths even at large repulsion. We thus have evidence 
to suggest, that the optical lattice barriers prevent the wave functions in each 
well from expanding to a comparable extent as in pure harmonic traps. We further 
found that the energy per particle is relatively insensitive to the number of
particles $N$ and variations in $a_c$ as compared to HS bosons in pure harmonic traps 
\cite{Dubois:01,Sakhel:02}, and that the tunneling amplitude increases 
with the reduction of the lattice spacing. In order to provide further support 
to our findings, we compare our VMC results with the corresponding VPI results. 

\hs Previous theoretical work on optical lattices is abundant, particularly 
the SF to MI transition 
\cite{Gerbier:05,Gygi:06,Li:06,Clark:06,Capello:01,Bach:04,Sansone:07,Yamashita:07}
and coherent matter waves \cite{Blackie:06,Bludov:06,Konotop:02,Rodriguez:06,Gerbier:05} 
have been investigated extensively. Other investigations included supersolidity 
\cite{Titvinidze:08}, multicomponent systems \cite{Zheng:05,Jin:05}, vortices 
\cite{Heiselberg:06}, solitons in a radially periodic lattice \cite{Baizakov:06}, 
and p-h excitations in MIs \cite{Gerbier:05}.

\hs Various techniques and methods have been used to investigate bosons, fermions, 
or mixtures of them in optical lattices: the Bose-Hubbard model 
\cite{Gygi:06,Sansone:07,Li:06,Barankov:07} as well as the Fermi-Hubbard model 
\cite{Rigol:04,Burovski:06} in conjunction with Monte Carlo techniques 
\cite{Gygi:06,Sansone:07,Rigol:04,Yamashita:07,Buonsante:04}, and particularly the 
Worm Algorithm \cite{Prokofev:98,Prokofev:99,Burovski:06} have been applied 
extensively. Other important techniques such as the Gross-Pitaevskii equation (GPE)
\cite{Baizakov:06,Huhtamaeki:07}, variational approaches 
\cite{Capello:01,Baizakov:06,Li:06}, density matrix renormalization group 
\cite{Rodriguez:06}, and path integral approaches \cite{Wouters:04} have also been 
used. Most of the above methods use a discrete space approach, whereas we evaluate 
the properties in continuous space.

\section{Method}

\hs We thus consider $N$ bosons on a combined harmonic optical cubic lattice 
(CHOCL). It consists of a 3D$^{\hbox{\it al}}$ simple cubic (SC) lattice of 
$N_L=3\times 3\times 3$ sites embedded in a tight external harmonic trap of 
frequency $\omega_{ho}$ and trap length $a_{ho}=\sqrt{\hbar/m\omega_{ho}}$, 
where $m$ is the mass of the bosons, and $\hbar$ is Planck's constant. The 
lattice spacing is given by $d\,=\,\pi/k$, where $k$ is the wave vector of 
the laser light. The bosons are modelled by hard spheres (HS) of diameter 
$a_c$ and their interactions $V_{int}(r)$ are represented by a hard-core 
potential of diameter $a_c$ given by

\begin{equation}
V_{int}(r)\,=\,\left\{ \begin{array}{r@{\quad:\quad}l} 
\infty & r \le a_c \\ 
0 & r > a_c \end{array} \right., \label{eq:Hard-sphere-potential}
\end{equation}

where $r$ is the distance between a pair of bosons. In the low-energy limit, 
the scattering between the bosons is purely s-wave with scattering length 
$a_s$. In this limit, $a_s$ equals $a_c$ \cite{Sakhel:02,Dubois:01} and the 
repulsive interactions between the bosons are modified by changing $a_c$.
Within this framework then, the tunneling amplitude $J$ and the rest of the 
properties are measured as functions of $a_c$. 

\hs To set the stage, then, we first define the Hamiltonian, and the 
trial wave function. Then the tunneling amplitude is measured by the overlap 
integral \Ip\ of two wave functions in neighboring wells centered at positions 
$\mathbf{R}_n$ and $\mathbf{R}_{n+1}$ measured from the center of the trap, 
where $n$ is an arbitrary integer. Here $n\equiv(pqr)$ are site indices with 
locations 
$\mathbf{R}_n\,=\,(p\mathbf{\hat{i}}+q\mathbf{\hat{j}}+r\mathbf{\hat{k}})$
in units of the lattice spacing $d$. By evaluating the overlap integral, the 
purpose is just to obtain a qualitative measure for the tunneling amplitude. 
Next the average onsite interaction energy $\langle U_n\rangle$ and the average 
occupation number per lattice site $\langle N_{pqr} \rangle$ at position 
$\mathbf{R}_n\equiv(pqr)$ are further defined. These are evaluated using 
VMC and VPI. We do not explain the VMC and VPI methods here as they can be 
found in the abundant literature 
\cite{Dubois:01,Cuervo:05,Sarsa:00,Kalos:86}.

\subsection{Hamiltonian}

\hs The Hamiltonian for our systems is given by

\begin{equation}
H\,=\,-\frac{\hbar^2}{2\,m}\sum_{i=1}^N\,\nabla_i^2\,+\,\sum_{i=1}^N\,
\left[V_{ho}(\mathbf{r}_i)\,+\,V_{opt}(\mathbf{r}_i)\right]+\sum_{i<j} V_{int}(r_{ij}),
\label{eq:Hamiltonian-harmonic-optical-lattice}
\end{equation}

where $V_{ho}(\mathbf{r}_i)\,=\,\frac{1}{2}m\omega_{ho}^2 r_i^2$ is an external 
harmonic trapping potential with $\mathbf{r}_i\,\equiv\,(x_i,\,y_i,\,z_i)$ representing 
a single particle position, $m$ the mass of the bosons, $V_{int}(r_{ij})$ with 
$r_{ij}=|\mathbf{r}_i-\mathbf{r}_j|$ the pair-interaction potential 
Eq.(\ref{eq:Hard-sphere-potential}) above, and 

\begin{equation}
V_{opt}(\mathbf{r}_i)\,=\,V_0\,\left[\sin^2(kx_i)\,+\,\sin^2(ky_i)\,+\,\sin^2(kz_i)\right], 
\label{eq:optical-potential}
\end{equation}

is the optical lattice potential \cite{Li:06} with $V_0$ the optical depth. Essentially, 
$V_{opt}(\mathbf{r}_i=\mathbf{R}_n)=0$ at the lattice-site positions $\mathbf{R}_n$ 
which are the locations of the potential-well minima of Eq.(\ref{eq:optical-potential}). 
That is, $V_{opt}(\mathbf{R}_n)=V_0[\sin^2(p\pi)+\sin^2(q\pi)+\sin^2(r\pi)]=0$ where 
$\mathbf{k}\cdot\mathbf{R}_n=\hbox{(integer)}\times \pi$. Thus the lattice-site positions 
are implied in $V_{opt}(\mathbf{r}_i)$ by its very construction. Experimentally, the optical 
lattice potential is obtained from a superposition of three pairs of mutually perpendicular, 
counterpropagating laser beams of intensity proportional to $V_0$. For a SC lattice,
$V_0$ must be the same for each direction. 
In this paper, we write energy and length in units of the trap, $\hbar\omega_{ho}$ 
and $a_{ho}\,=\,\sqrt{\hbar/m\omega_{ho}}$, respectively.

\subsection{Trial wave function}

\hs The many-body trial wave function is given by
\vspace{-0.1cm}
\begin{equation}
\Psi(\{\mathbf{r}\}, \{\mathbf{R}\})\,=\,\prod_{i=1}^N\,\exp(-\alpha\,r_i^2)\,
\psi(\mathbf{r}_i,\{\mathbf{R}\})\,\prod_{i < j} f(|\mathbf{r}_i-\mathbf{r}_j|), 
\label{eq:Trial-wave-function}
\end{equation}
with $\psi(\mathbf{r}_i,\{\mathbf{R}\})$ a Wannier-like function defined as

\begin{equation}
\psi(\mathbf{r}_i,\{\mathbf{R}\})\,=\,\sum_{n=0}^{N_L} \phi(\mathbf{r}_i, \mathbf{R}_n), 
\label{eq:Wannier-like-function}
\end{equation}
where $\{\mathbf{r}\}\,\equiv\,\{\mathbf{r}_1,\mathbf{r}_2,\cdots,\mathbf{r}_{N}\}$ 
is a set of spatial vectors describing the positions of the bosons from the 
center of the trap, 
$\{\mathbf{R}\}\,\equiv\,\{\mathbf{R}_1,\mathbf{R}_2,\cdots,\mathbf{R}_{N_L}\}$
is a set of vectors describing the positions of the lattice potential minima on 
the $3\times 3\times 3$ cubic lattice cage considered, and $\alpha$ is a variational 
parameter signalling the strength of the external harmonic confinement. Essentially, 
$\alpha$ is the inverse overall width of the total wave function of the system in the 
CHOCL. Further, $\alpha$ controls the volume of the external harmonic confinement 
and, therefore, the number of lattice sites to be occupied away from the center of 
the trap. Eq.(\ref{eq:Wannier-like-function}) is constructed in a manner similar to 
that of Jin \ea\ \cite{Jin:05} in that we sum over a number of localized single-particle 
wave functions $\phi(\mathbf{r}_i,\mathbf{R}_n)$ centered in the optical lattice wells, 
each at $\mathbf{R}_n$. $\phi(\mathbf{r}_i,\mathbf{R}_n)$ is constructed similarly 
to the wave function used by Li \ea\ \cite{Li:06} and is given by:
\vspace{-0.5cm}

\begin{eqnarray}
\phi(\mathbf{r}_i,\mathbf{R}_n)\,&=&\,\exp[-\beta(\mathbf{r}_i-\mathbf{R}_n)^2]\times\,\nonumber\\
&&\left[1+\gamma(x_i-X_n)^2-\sigma(x_i-X_n)^4\right]\times \nonumber\\
&&\left[1+\gamma(y_i-Y_n)^2-\sigma(y_i-Y_n)^4\right]\times \nonumber\\
&&\left[1+\gamma(z_i-Z_n)^2-\sigma(z_i-Z_n)^4\right], \label{eq:phi_similar_to_Li}
\end{eqnarray}

where $\beta$, $\gamma$, and $\sigma$ are further variational parameters in 
addition to $\alpha$. The onsite repulsion is partly controlled by the 
local density $|\phi(\mathbf{r}_i,\mathbf{R}_n)|^2$ via the variational parameter 
$\beta$ in Eq.\,(\ref{eq:phi_similar_to_Li}), which confines the particles at 
each lattice site ($n$). The interactions in the trial wave function are taken 
into account by the usual HS Jastrow function \cite{Sakhel:02,Dubois:01}:

\begin{equation}
f(r_{ij})\,=\,\left\{\begin{array}{r@{\quad : \quad}l} 
0 & r_{ij} \le a_c \\
1-\frac{a_c}{r_{ij}} & r_{ij} > a_c \\
\end{array}\right., \label{eq:Hard-sphere-Jastrow}
\end{equation}

where $r_{ij}$ is the distance between a pair of bosons.

\hs The VMC wave function (\ref{eq:Trial-wave-function}) is optimized by 
minimizing the average energy with Powell's technique \cite{Press:99} employed 
in a previous publication \cite{Sakhel:08}. The ground state configuration is 
achieved when all the particles are distributed symmetrically in the spherical 
coordination shells around the center of the trap, and the overall boson-boson 
repulsion is minimal. Once the trial function Eq.(\ref{eq:Trial-wave-function}) 
has been optimized, it is plugged into the VPI code as a starting wave function. 
The same properties are then evaluated as those using VMC.

\subsection{Tunneling}

\hs Since we consider in this work only the strongly interacting regime, 
single-particle tunneling is very much suppressed in our systems. Hence, the 
atom-pair tunneling is dominant and is included in the overlap of the wave 
functions in neighboring wells. A qualitative measure for the overlap of the wave 
functions is evaluated, for simplicity, from trial-density functions fitted to 
slices of the integrated optical densities along one axis. For this purpose, 
the trial-density functions are similar to the one-dimensional version of 
Eq.(\ref{eq:phi_similar_to_Li}) with an additional amplitude factor $A_n$, that is

\begin{figure}[t!]
\includegraphics*[width=8.5cm,viewport=180 544 537 790]{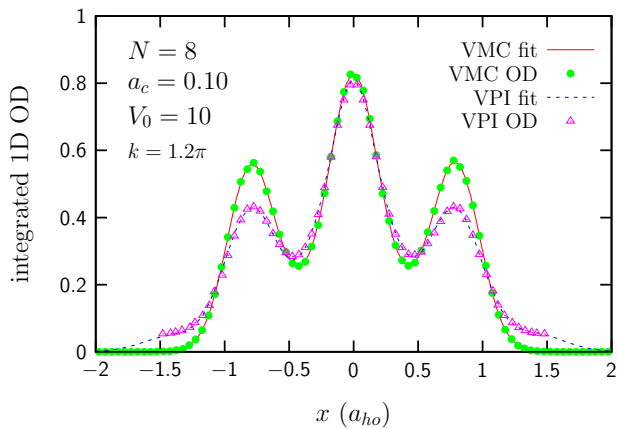}
\caption{Fits using the forms (\ref{eq:fitting-function}) and
(\ref{eq:total-fitting-function}) to 1D VMC 
and VPI integrated optical density slices along the x-axis for a system of 
$N=8$, $V_0=10$, and $k=1.2\pi$. The sum of squares for both fits is 
$3.41\times 10^{-4}$ for VMC and $7.41\times 10^{-4}$ for VPI, 
respectively.} 
\label{fig:plot.fittodensity.VMC.VPI.ODx.N8.ac0.10.B10.k1.2}
\end{figure}

\begin{eqnarray}
&&\phi_{fit}(x,X_n)^2\,=\,A_n\,\exp[-\beta_n(x-X_n)^2]\,\times\nonumber\\
&&\left|1+\gamma_n(x-X_n)^2+\sigma_n(x-X_n)^4 \right|. \label{eq:fitting-function}
\end{eqnarray}

The absolute value is considered to make sure we do not run into negative
densities. An example of such a fit is shown in 
Fig.\ft\ref{fig:plot.fittodensity.VMC.VPI.ODx.N8.ac0.10.B10.k1.2} below.
Here, the fitting function is: 

\begin{equation}
F(x;X_1,X_2,X_3)=\sum_{n=1}^3 \phi_{fit}(x,X_n)^2 \label{eq:total-fitting-function}
\end{equation}

having peaks centered at three potential minima $X_1$ (left), $X_2$ (center), 
and $X_3$ (right) and with three sets of fitting parameters {$A_n$, $\beta_n$, 
$\gamma_n$, $\sigma_n$, and the latter $X_n$}, where $n$ runs from 1 to 3. 
In fact, the central position $X_2$ always turns out to be exactly zero as 
required. The sum of squares 

\begin{equation}
\chi^2\,=\,\sum_{i=1}^P\left\{\sum_{n=1}^3 
\left(|\phi_{fit}(x_i,X_n)|^2-|\phi_{MC}(x_i,X_n)|^2\right)\right\}^2 
\label{eq:sum-of-squares}
\end{equation}

is minimized with respect to the above sets of fitting parameters. Here
$\chi^2$ is a sum over all $P$ data points of the one-dimensional integrated 
MC optical density [$\sum_{n=1}^3|\phi_{MC}(x_i,X_n)|^2$]. By this minimization,
we get values of $\chi^2 < 10^{-3}$ indicating a good fit. After optimization,
the overlap integral

\begin{eqnarray}
&&I_{overlap}\,=\,\int_{-\infty}^{+\infty} \phi_{fit}(x,X_1)\,\phi_{fit}(x,X_2)\,dx + 
\nonumber\\
&&\int_{-\infty}^{+\infty}\,\phi_{fit}(x,X_2)\,\phi_{fit}(x,X_3)\,dx,
\label{eq:overlap-integral}
\end{eqnarray}

is then evaluated using a simple elementary numerical technique. 

\hs We must emphasize, that it is hard to describe the single-particle 
tunneling amplitude for strongly interacting systems using the well-known 
exchange integral \cite{Li:06},

\begin{equation}
J_n\,=\,\displaystyle\int d^3 r \phi_n(\mathbf{r})
\left[-\frac{\hbar^2}{2m}\nabla_i^2+V(\mathbf{r})\right]\phi_{n+1}(\mathbf{r}), 
\label{eq:J_n_general}
\end{equation}

where $\phi_n$ and $\phi_{n+1}$ are single-particle wave functions at sites
$n$ and $n+1$, and $V(\mathbf{r})$ a single-particle potential. This is
because the interactions as described by the Jastrow factor 
[Eq.(\ref{eq:Hard-sphere-Jastrow})] are not included in Eq.(\ref{eq:J_n_general}).
In fact, it is anticipated that the single-particle wave function narrows as
the HS repulsion rises, counteracting the effects of the broadening due to
the Jastrow factor as the number of particles is increased beyond a certain
limit. 

\subsection{Bosons-optical lattice overlap}\label{sec:disorder-integral}

\hs A measure for the extent of the overlap between 
$\phi(\mathbf{r}_i, \mathbf{R}_n)$ and $V_{opt}(\mathbf{r}_i)$ for all particles 
$i=1$ to $N$ and all lattice sites $n=1$ to $N_L$ is given by

\begin{equation}
\langle I_{BOSON-OL} \rangle\,=\,\sum_{n=1}^{N_L} 
\displaystyle \int_{\langle MC \rangle} 
e^{-\alpha r^2}\phi(\mathbf{r}, \mathbf{R}_n)\,V_{opt}
(\mathbf{r})\,d^3 r, \label{eq:disorder-integral}
\end{equation}

where $\int_{\langle MC \rangle}$ stands for a MC configurational
integral defined in Sec.\ref{sec:numerics} below, and $OL$ stands for optical 
lattice. Further $\langle I_{BOSON-OL}\rangle$ is an indirect measure of 
the total tunneling amplitude of the system into the optical lattice 
potential barriers.

\subsection{Discrete onsite interaction energies}

\hs In order to explore the single-particle response to the change in the
HS repulsions, we evaluate the onsite interaction energies at individual
lattice sites. The onsite interaction energy for a single particle at 
each lattice site $(n)$ is evaluated using \cite{Li:06,Morsch:06}

\begin{equation}
\langle U_{n}\rangle\,=\,g\displaystyle
\int_{\langle MC\rangle} e^{-4\alpha r^2}
|\phi(\mathbf{r},\mathbf{R}_n)|^4\,d^3 r,
\label{eq:discrete-onsite-interaction}
\end{equation}

where $g=4\pi\hbar^2 a_c/m$ with $a_s$ replaced by $a_c$. Here the effects 
of the external trap are included via the factor $[\exp(-\alpha r^2)]^4$.

\subsection{Occupancy of Lattice Sites}\label{sec:occupancy-of-sites}

\hs Another single-particle response is the atom-number occupancy of each
lattice site. The average number of particles occupying a particular lattice 
site $(n)$ is evaluated via

\begin{equation}
\langle N_n \rangle\,=\,N\,\int_{\langle MC \rangle} e^{-2\alpha r^2}
|\phi(\mathbf{r},\mathbf{R}_n)|^2\,d^3 r. 
\label{eq:occupancy}
\end{equation}

An alternative way to determine the occupancy of each lattice site, is to 
divide the confining volume into equivalent cubic bins whose corners are 
sets of four lattice sites $\mathbf{R}_n$ and count the number of particles 
collected in each bin. We used this counting method to evaluate the VPI
$\langle N_n \rangle$, whereas (\ref{eq:occupancy}) to obtain the VMC
$\langle N_n \rangle$.

\subsection{Correlations between the bosons}

\hs The pair correlation function $g(r)$ is evaluated using VPI by 
binning pairs of particles in each $r$. The idea is to provide a measure 
for the strength of the boson-boson correlations and its dependency on 
the repulsive interactions in a tightly confining environment.

\subsection{Numerics}\label{sec:numerics}

\hs The integrals (\ref{eq:disorder-integral}), 
(\ref{eq:discrete-onsite-interaction}) and (\ref{eq:occupancy}) 
are evaluated using standard Monte Carlo integration \cite{Kalos:86}:

\begin{equation}
\displaystyle\int_{\langle MC\rangle} O(\mathbf{r}_i) 
d\mathbf{r}_i\,=\,\frac{1}{N}\sum_{i=1}^N\frac{1}{M}\sum_{c=1}^M 
\frac{O(\mathbf{r}_{ci})}{e^{-2\alpha r_{ci}^2} 
\sum_{n=1}^{N_L} |\phi(\mathbf{r}_{ci},\mathbf{R}_n)|^2}, 
\label{eq:Monte-Carlo-Integration}
\end{equation}

where $c$ is a configurational index, $i$ a particle index, and $M$ is 
the total number of VMC configurations. We further sum over all $N$
particles and divide by $N$ to get the average. The denominator is a 
weight we used for all MC integrals. In VPI we apply the same integration
technique, except with $M$ taken as the total number of time slices used 
in the path integral. Each time slice contains one configuration of $N$ 
particle positions.

\subsection{Computational complexities}\label{sec:comp-complexity}

\hs As much as VPI is an accurate method for the present determination of
the properties of lattice bosons, using a number of particles exceeding 
$N=8$, this method begins to constitute a heavy-computational and CPU-time 
consuming technique. Particularly, for large optical potential barriers 
$V_0\ge 10\,\hbar\omega_{ho}$, the bosons take a very long CPU time to tunnel 
through the barriers in order to achieve a symmetric optical density 
distribution inside the whole trap. We therefore limited the number of 
particles to $N=8$ when using VPI. On the other hand, the less accurate 
VMC method allows the bosons to diffuse quickly through the barriers providing 
a computationally cheap and fast technique to investigate qualitatively the 
properties of bosons in optical lattices. Our aim is, however, not to present 
methods but to concentrate on the physics of quantum tunneling.

\section{Results}\label{sec:results}

\hs In what follows, we present the results of our calculations. We 
particularly focus on the effect of interactions on the tunneling amplitude 
$\propto$ \Ip. We further explore the effect of $a_c$ on the onsite 
interaction energies $\langle U_n \rangle$, energy per particle 
$\langle E\rangle/N$, the occupancy of lattice sites $\langle N_n \rangle$, 
the correlations $g(r)$, and the optical density. As mentioned before, we 
compare our VMC results with corresponding VPI results.

\subsection{Tunneling}

\begin{figure}[t!]
\includegraphics*[width=8.5cm,viewport=177 522 541 775,clip]{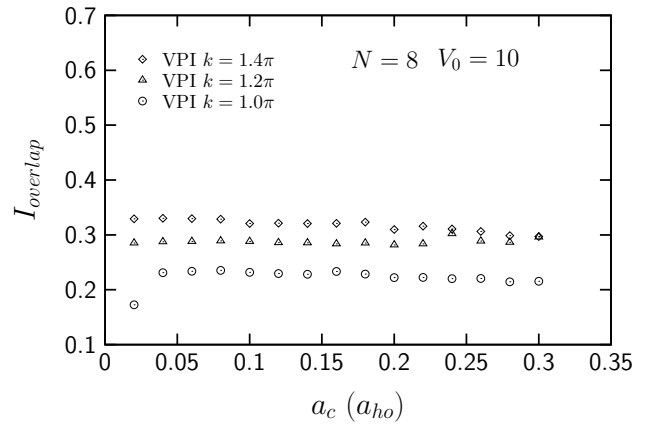}
\caption{Tunneling amplitude $\propto$ \Ip\ [Eq.(\ref{eq:overlap-integral})] vs 
the hard-sphere (HS) diameter $a_c$, obtained from fits to VPI optical
densities. The system is a HS Bose gas of $N=8$ particles in a 
$3\times 3\times 3$ cubic optical lattice of depth $V_0=10$ embedded in a 
tight external harmonic trap of trapping frequency $\omega_{ho}$. 
Open circles: $k=\pi$, open triangles: $k=1.2\pi$, and open diamonds: 
$k=1.4\pi$. Energies and lengths are in units of the trap.} 
\label{fig:plot.ovrlp.vpi.N8.B10.severalk}
\end{figure}

\hs \Ip\ is now displayed as a function of $a_c$. 
Fig.\ft\ref{fig:plot.ovrlp.vpi.N8.B10.severalk} displays the VPI results for 
systems of $N=8$ and $V_0=10$ with $k=\pi$ (open circles), $k=1.2\pi$ (open 
triangles) and $k=1.4\pi$ (open diamonds). For these values of $N$ and $V_0$, 
\Ip\ shows a very weak response to the effects of increasing $a_c$, since
it seems to remain practically constant. As $k$ is increased, \Ip\ rises 
indicating a profound effect of the lattice spacing $d=\pi/k$ on the tunneling 
amplitude. 

\begin{figure}[t!]
\includegraphics*[width=8.5cm,viewport=177 521 543 779,clip]{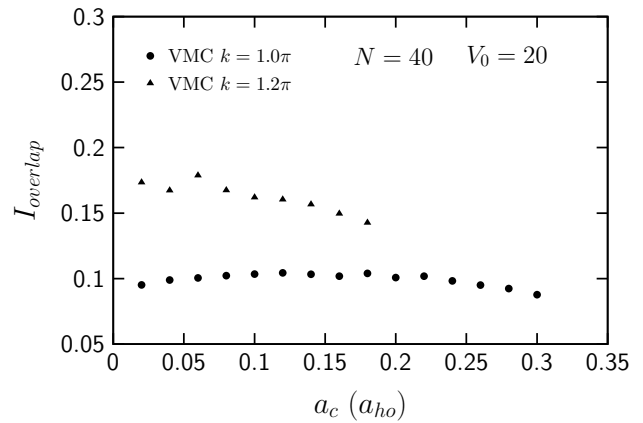}
\caption{As in Fig.\ft\ref{fig:plot.ovrlp.vpi.N8.B10.severalk}; but 
using VMC for $N=40$, $V_0=20$ and the indicated values of $k$.} 
\label{fig:plot.ovrlp.vmc.N40.B20.severalk}
\end{figure}

\begin{figure}[t!]
\includegraphics*[width=8.5cm,viewport=175 521 534 777,clip]{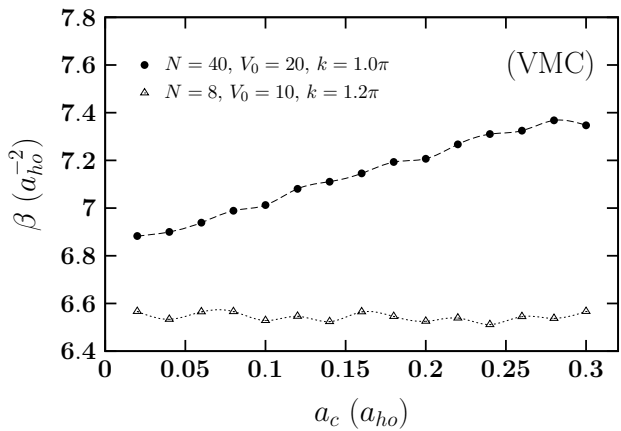}
\caption{VMC parameter $\beta$ of the trial wave function 
[Eq.(\ref{eq:phi_similar_to_Li})] as a function of the hard-sphere diameter 
$a_c$ for two HS Bose gas systems in a combined harmonic simple cubic 
optical lattice of $3\times 3\times 3$ sites. Solid circles: HS Bose gas 
of Fig.\ft\ref{fig:plot.ovrlp.vpi.N8.B10.severalk} at $k=1.2\pi$. Open 
triangles: HS Bose gas of 
Fig.\ft\ref{fig:plot.ovrlp.vmc.N40.B20.severalk} at $k=\pi$.}
\label{fig:plot.width.vs.ac.several.systems}
\end{figure}

\begin{figure}[t!]
\includegraphics*[width=8.5cm,viewport = 207 456 574 709,clip]{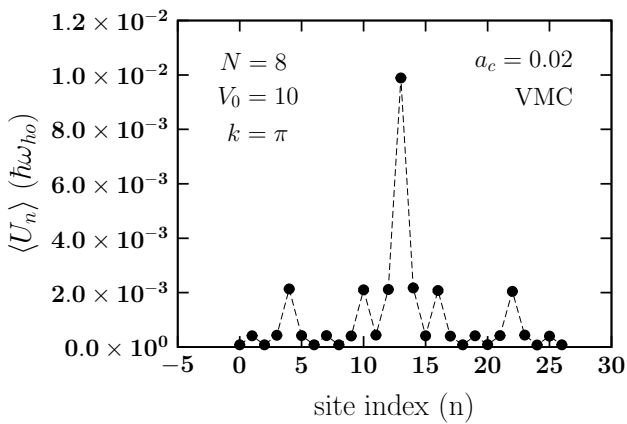}
\caption{Average VMC onsite repulsive interaction energy $\langle U_n \rangle$ at 
each lattice site $n$ for a HS Bose gas of $N=8$ particles, optical lattice 
depth $V_0$, and HS diameter $a_c=0.02$ in a $3\times 3\times 3$ cubic 
optical lattice embedded in a tight external harmonic trap of trapping 
frequency $\omega_{ho}$. Energies and lengths are in units of the trap.} 
\label{fig:plot.U.vs.sites.V10.N8.ac0.02.L27.k1.0}
\end{figure}

\begin{figure}[t!]
\includegraphics*[width=9.0cm,viewport=203 456 573 711,clip]{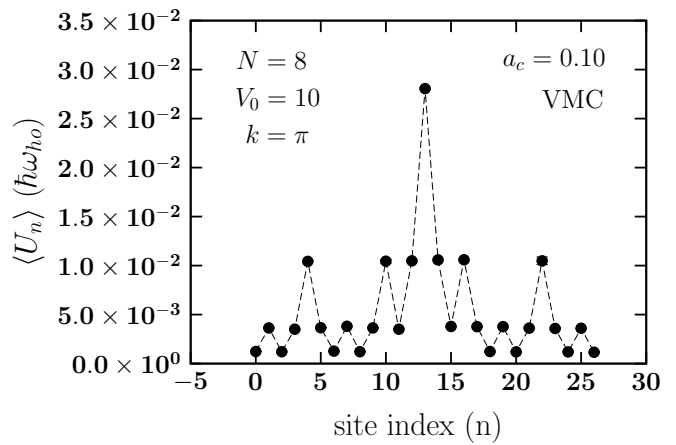}
\caption{As in Fig.\ft\ref{fig:plot.U.vs.sites.V10.N8.ac0.02.L27.k1.0}; 
but for $a_c=0.10$.} \label{fig:plot.U.vs.sites.V10.N8.ac0.10.L27.k1.0}
\end{figure}

\hs The case for $N=40$ and $V_0=20$ is shown in 
Fig.\ft\ref{fig:plot.ovrlp.vmc.N40.B20.severalk} for $k-$values of $\pi$ 
and $1.2\pi$ (solid circles and triangles, respectively). For $k=1.2\pi$, 
\Ip\ decreases overall with increasing $a_c$, whereas for $k=1.0\pi$ it rises
somewhat up to $a_c=0.18$ and then begins to drop. However, the response
is still weak and insignificant. Hence, a slightly stronger response is 
obtained for the tunneling amplitude when using a larger $N$. Further,
the effect of changing $d=\pi/k$ on $I_{overlap}$ is much more pronounced
than changing $a_c$. Signals for a stronger response of our systems to 
the rise of $a_c$ are found in the single-particle properties such as the 
width of the single-particle wave function and the onsite interaction energy.

\subsection{Width of the single-particle wave function in each potential well}

\hs As already indicated in the Method section, the width of the 
single-particle wave function $\phi(\mathbf{r},\mathbf{R}_n)$ 
[Eq.(\ref{eq:phi_similar_to_Li})] in each well, is expected to decrease 
with a rise of the HS repulsion in order 
to counteract the effects of a broadening due to the Jastrow function 
for a larger number of particles. The width, or better the inverse of it, 
is described by the VMC parameter $\beta$ in Eq.(\ref{eq:phi_similar_to_Li}). 
In Fig.\ft\ref{fig:plot.width.vs.ac.several.systems}, the VMC parameter 
$\beta$ of the local wave function, $\phi(\mathbf{r},\mathbf{R}_n)$, is 
displayed as a function of $a_c$. The open triangles display $\beta$ for 
the HS Bose gas of Fig.\ft\ref{fig:plot.ovrlp.vpi.N8.B10.severalk} at 
$k=1.2\pi$; solid circles: HS Bose gas of 
Fig.\ft\ref{fig:plot.ovrlp.vmc.N40.B20.severalk} at $k=\pi$. It is found 
that this width $(\propto 1/\sqrt{\beta})$ decreases with the rise in 
the repulsion for $N=40$ and remains almost constant for $N=8$.

\subsection{Onsite interaction energies}

\hs Discrete distributions of $\langle U_n\rangle$ over all 27 lattice sites, 
for systems of $N=8$, $V_0=10$, $k=\pi$, and two cases $a_c=0.02$ and 0.10, are 
displayed in Figs.\ft\ref{fig:plot.U.vs.sites.V10.N8.ac0.02.L27.k1.0} and 
\ft\ref{fig:plot.U.vs.sites.V10.N8.ac0.10.L27.k1.0}, respectively. The abcissa 
display the site index $n$. All $\langle U_n\rangle$ which have the same magnitude 
within a small margin of error, belong to the same set of nearest neighbors from 
the central lattice site. That is, moving vertically from the tip of the discrete 
distribution down to the bottom, we move from the center to the first, second, and 
third set of nearest neighbors, respectively. The distribution is uniform around the 
trap center. The effects of the external trap are manifested in the magnitude of 
$\langle U_{n} \rangle$ which is highest at the center and declines towards the 
edges of the trap due to the factor $\exp(-4\alpha r^2)$ in 
Eq.(\ref{eq:discrete-onsite-interaction}). Obviously, a large percentage of the 
particles is concentrated at the central lattice site. (It is anticipated then, 
that if the strength of the external harmonic trap is increased further, this causes 
a drop in the density $\sim \exp(-2\alpha r^2)\,|\phi(\mathbf{r},\mathbf{R}_n)|^2$ 
at the lattice sites towards the edges of the trap and a certain rise at the
center of the trap.) Eventually, all the 
bosons will occupy the central lattice site and only if their HS diameter
is of a magnitude that allows all of them them to fit inside one lattice-site 
volume $\sim 4\pi(d/2)^3/3$. Thus, a large amount of repulsive potential energy 
would be stored in the trap center. The patterns of 
Figs.\ft\ref{fig:plot.U.vs.sites.V10.N8.ac0.02.L27.k1.0} and 
\ft\ref{fig:plot.U.vs.sites.V10.N8.ac0.10.L27.k1.0}
show also that the bosons minimize their potential energy arising from the external 
harmonic trap by maximizing their occupancy of the trap center, and minimizing it 
towards the edges of the trap.

\begin{figure}[t!]
\includegraphics*[width=8.5cm,viewport = 172 455 534 711,clip]{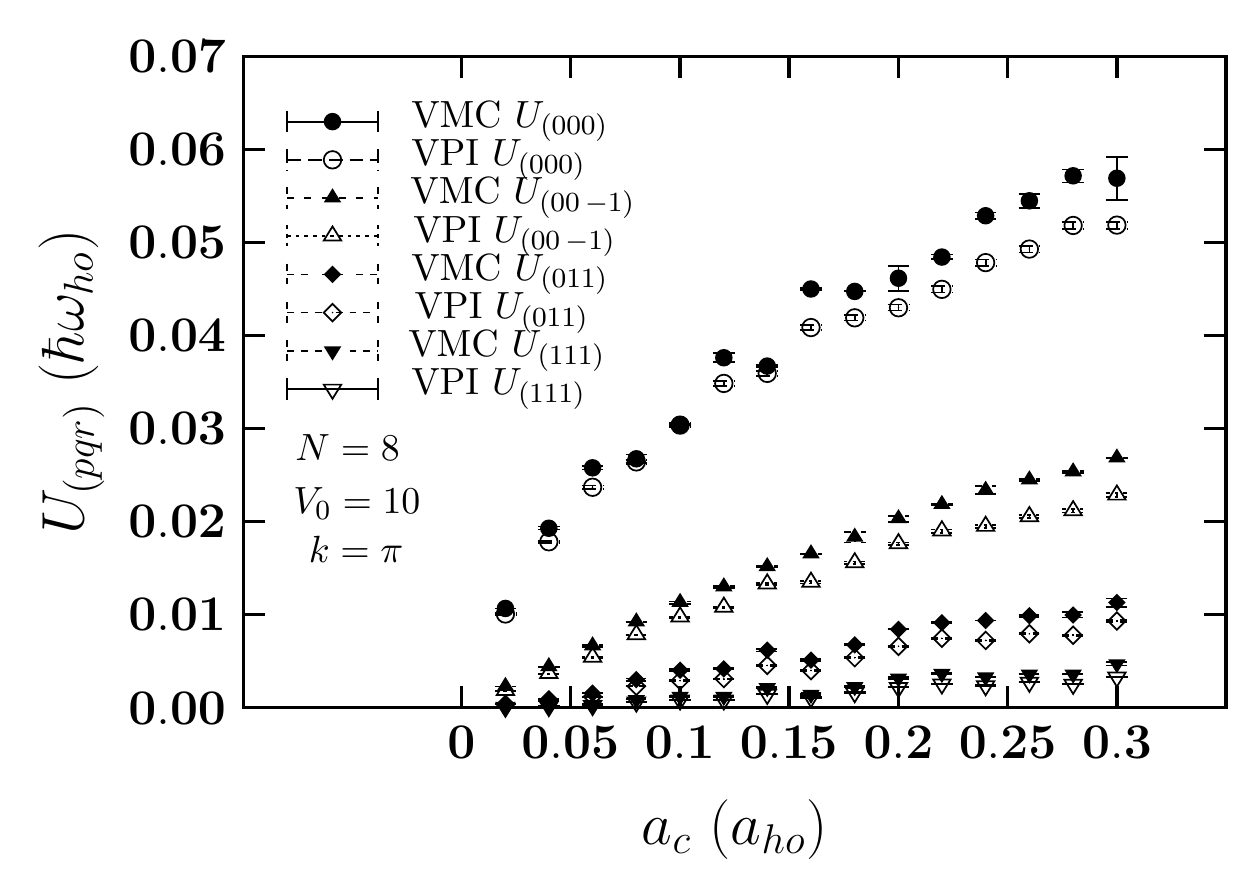}
\caption{Average onsite repulsive interaction energy $\langle U_{(pqr)} 
\rangle$ vs the hard-sphere (HS) diameter $a_c$ at four lattice sites 
representative of the whole lattice. The system is the HS Bose gas mentioned
in Fig.\ft\ref{fig:plot.ovrlp.vpi.N8.B10.severalk} at $k=\pi$. Solid and open 
circles: VMC and VPI results for $\langle U_{(000)}\rangle$. Solid and open 
triangles: likewise for $(00\,-1)$. Solid and open diamonds: for
$(011)$. Solid and open inverted triangles: for $(111)$. } 
\label{fig:plot.U.for.4.sites.vs.ac.V10.N8.k1.0.L27}
\end{figure}

\begin{figure}[t!]
\includegraphics*[width=8.5cm,viewport = 172 455 534 711,clip]{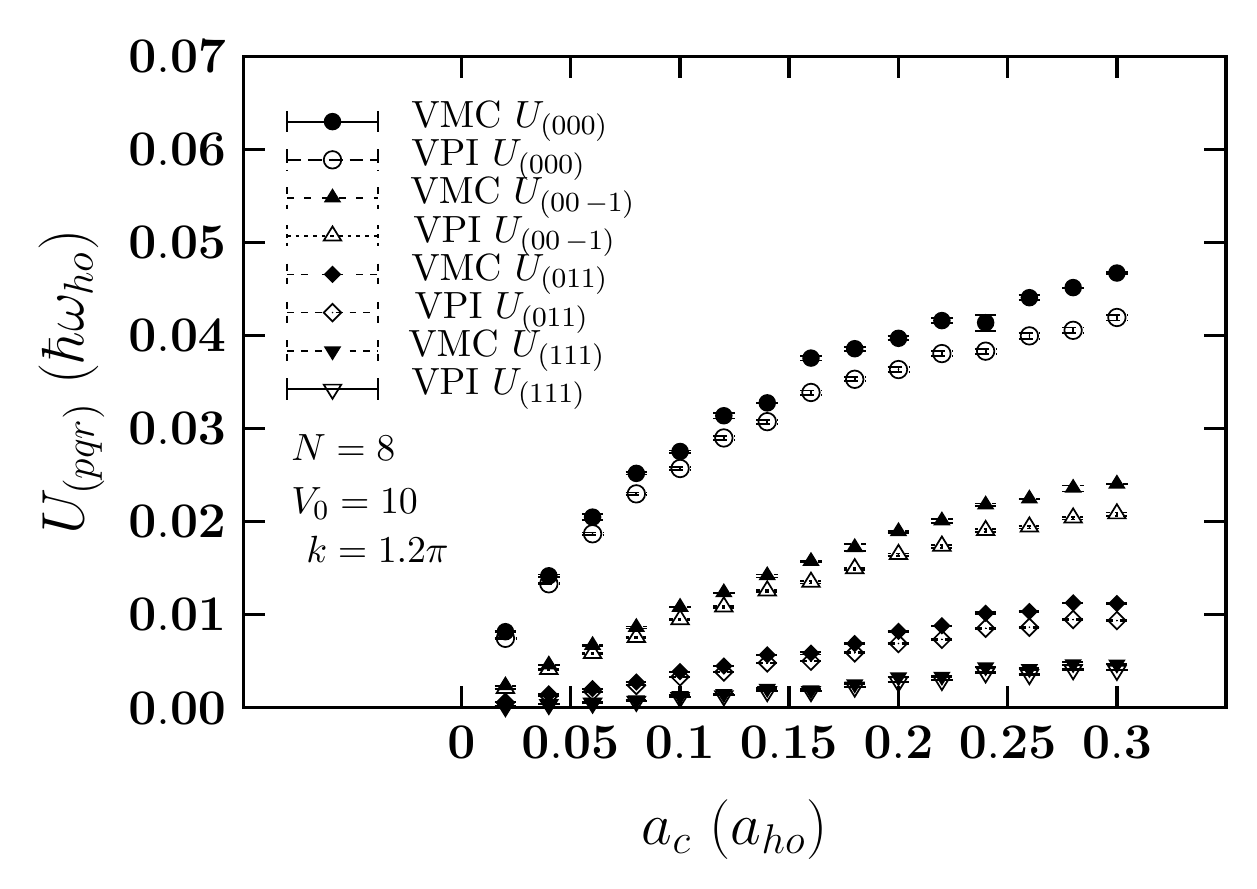}
\caption{As in Fig.\ft\ref{fig:plot.U.for.4.sites.vs.ac.V10.N8.k1.0.L27}; 
but for $k=1.2\pi$.} \label{fig:plot.U.for.4.sites.vs.ac.V10.N8.k1.2.L27}
\end{figure}

\hs In addition, $\langle U_n \rangle$ is displayed as a function of $a_c$ in 
Figs.\ft\ref{fig:plot.U.for.4.sites.vs.ac.V10.N8.k1.0.L27} and 
\ref{fig:plot.U.for.4.sites.vs.ac.V10.N8.k1.2.L27} for the systems of 
Fig.\ft\ref{fig:plot.ovrlp.vpi.N8.B10.severalk} 
at $k=\pi$ and $1.2\pi$, respectively, and for four lattice sites: 
$\mathbf{R}_n\equiv(000)$, $(00-1)$, $(111)$, and $(011)$, representative of the 
whole lattice. In general, $\langle U_{(pqr)} \rangle$ rises with $a_c$ signaling 
a rise in the repulsive energy of the particles in each well. The rate of growth
of $\langle U_{(pqr)}\rangle$ is largest for the trap center and declines towards
the edges of the trap.

\begin{figure}[t!]
\includegraphics*[width=8.5cm,viewport=178 456 543 710,clip]{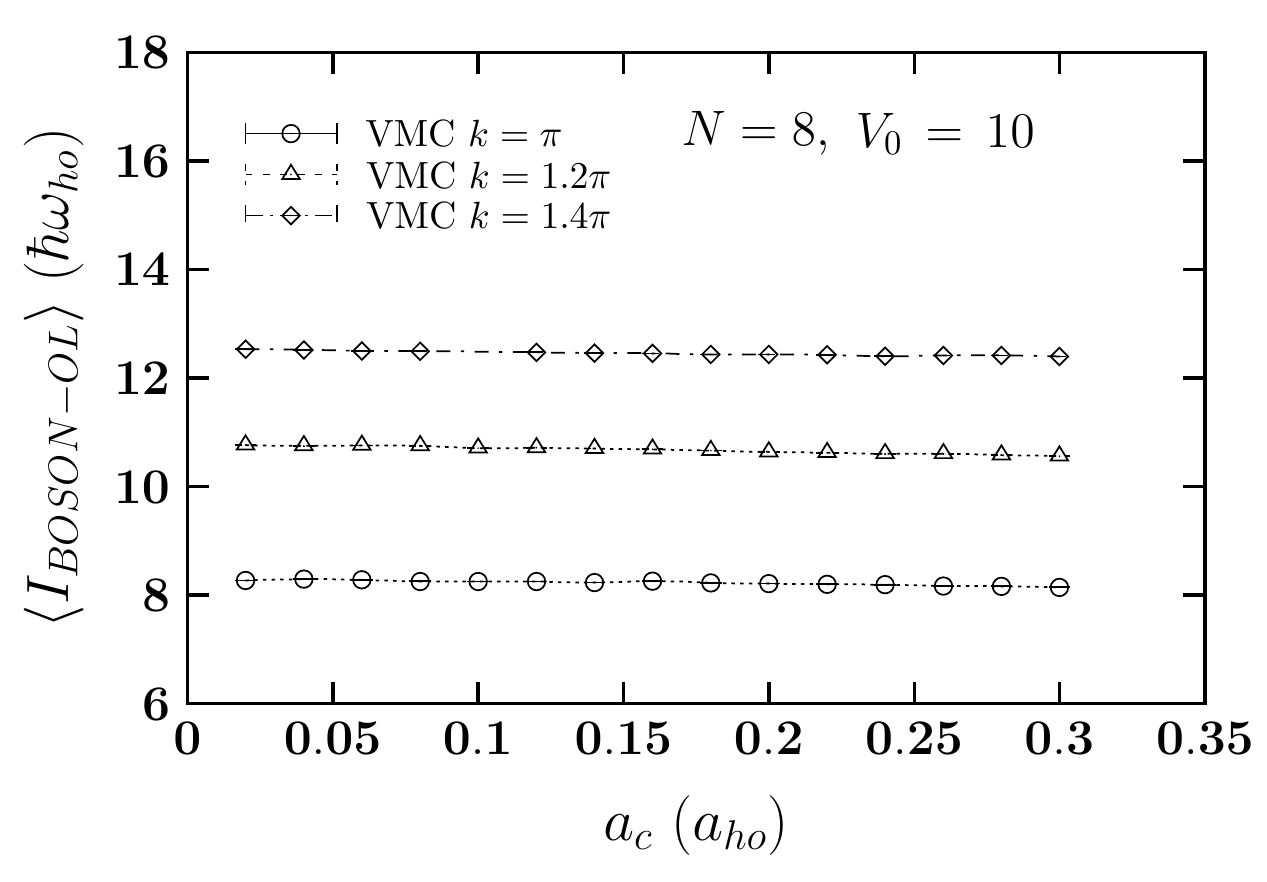}
\caption{VMC disorder integral [Eq.(\ref{eq:disorder-integral})] vs the 
hard-sphere diameter $a_c$ for the systems in 
Fig.\ft\ref{fig:plot.ovrlp.vpi.N8.B10.severalk}.} 
\label{fig:plot.Hdisorder.vs.ac.V10.N8.L27.severalk}
\end{figure}

\begin{figure}[t!]
\includegraphics*[width=8.5cm,viewport=176 456 538 710,clip]{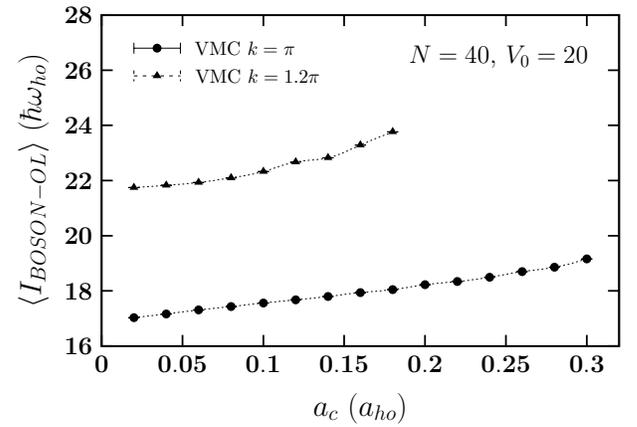}
\caption{VMC disorder integral [Eq.(\ref{eq:disorder-integral})] vs the 
hard-sphere diameter $a_c$ for the systems in 
Fig.\ft\ref{fig:plot.ovrlp.vmc.N40.B20.severalk}.} 
\label{fig:plot.Hdisorder.vs.ac.V20.N40.L27.severalk}
\end{figure}

\subsection{Boson-optical lattice overlap}

\hs In this section, we evaluate $\langle I_{BOSONS-OL} \rangle$ 
[Eq.\ft(\ref{eq:disorder-integral})] which measures the overlap of all the single 
particle wave functions in the wells with the periodic optical lattice 
potential. In essence, it measures also the {\it total} tunneling amplitude 
of the particles into the optical lattice potential barriers. 
Figs.\ft\ref{fig:plot.Hdisorder.vs.ac.V10.N8.L27.severalk} and
\ref{fig:plot.Hdisorder.vs.ac.V20.N40.L27.severalk} display \Ibol\ vs. $a_c$
for the systems of Figs.\ft\ref{fig:plot.ovrlp.vpi.N8.B10.severalk} and 
\ref{fig:plot.ovrlp.vmc.N40.B20.severalk}, respectively. One can see that
\Ibol\ is practically invariant for $N=8$, but the system with $N=40$
reveals a slightly stronger response to changes in $a_c$. For example 
for $N=8$ and $k=\pi$, the change in \Ibol\ from $a_c=0.02$ to 0.3 is only 
by $-1.45\%$. The other changes for $N=8$ are of the same order of magnitude. 
This in turn indicates, that the width of each localized wave function for these 
systems is practically invariant with the rise in repulsion. This may come as a 
surprise, since it is known \cite{Sakhel:02,Dubois:01} that at very large HS 
diameters such as the ones used here, the wave function ought to become very 
broad indeed. It seems that the localization effect of the optical lattice 
potential overwhelms the repulsive forces tending to broaden the wave 
function in each well.

\hs One further observes that \Ibol\ increases with $k$ in agreement 
with the results of Figs.\ft\ref{fig:plot.ovrlp.vpi.N8.B10.severalk} and 
\ref{fig:plot.ovrlp.vmc.N40.B20.severalk}, where \Ip\ rises with increasing 
$k$. 

\hs Further, we note that in 
Fig.\ft\ref{fig:plot.Hdisorder.vs.ac.V20.N40.L27.severalk} \Ibol\ rises 
notably with $a_c$ indicating a rise in the total tunneling amplitude 
into the optical potential barriers. We were unable to obtain values of 
\Ibol\ for $k=1.2\pi$ beyond $a_c=0.18$ as we could not find VMC energy 
minima for these systems. We must emphasize that the overlap of the bosons 
with the optical lattice potential is a different measure than \Ip, which 
measures the overlap of two wave functions in neighboring wells.

\subsection{Occupancy of lattice sites}\label{sec:occupancy-results}

\begin{figure}[t!]
\includegraphics*[width=8.5cm,viewport=62 225 313 709,clip]{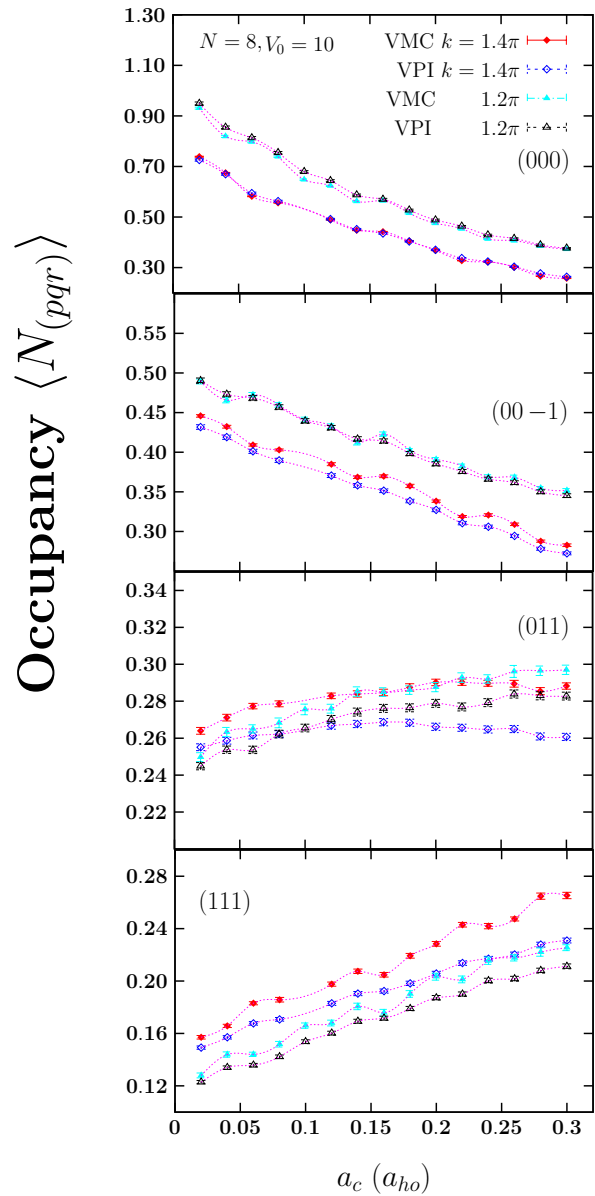}
\caption{(Color online) Average occupancy $\langle N_{(pqr)}\rangle$ vs the 
hard-sphere diameter $a_c$ at the four lattice sites $(pqr)$ indicated, 
which are representative of the whole optical lattice. The system is the HS 
Bose gas of Fig.\ft\ref{fig:plot.ovrlp.vpi.N8.B10.severalk}. 
Solid and open diamonds: VMC and VPI results for $k=1.4\pi$, respectively. Solid and open 
triangles: likewise, but for $k=1.2\pi$. From top to bottom frame: 
$\mathbf{R}_{13}\equiv(000)$ is the center, $\mathbf{R}_{12}\equiv(00\,-1)$ the first, 
$\mathbf{R}_{17}\equiv(011)$ the second, and $\mathbf{R}_{26}\equiv(111)$ the third 
nearest neighbor to the center, respectively. Site indices $n$ are according to
Fig.\ft\ref{fig:plot.U.vs.sites.V10.N8.ac0.02.L27.k1.0}.} 
\label{fig:plot.occupN.vs.ac.V10.N8.L27.severalk.figurestack}
\end{figure}

\hs Figure\ft\ref{fig:plot.occupN.vs.ac.V10.N8.L27.severalk.figurestack} 
displays the average occupancy of four lattice sites $\langle N_{(pqr)} \rangle$ 
for systems of Fig.\ft\ref{fig:plot.ovrlp.vpi.N8.B10.severalk}: 
$(pqr)\equiv(000)$, $(00\,-1)$, $(011)$, and $(111)$. Each frame is labelled 
by the corresponding lattice site. The solid and open diamonds display VMC 
and VPI results for $k=1.4\pi$, whereas the solid and open triangles those 
for $k=1.2\pi$, respectively. The VMC results have been obtained by 
Eq.(\ref{eq:occupancy}), whereas the VPI results from simple counting of the 
bosons in cubic bins. In frames (000) and $(00\,-1)$, the occupancy decreases 
with $a_c$, whereas in frames (011) and (111) it rises. Further, the occupancy 
is higher for a lower $k$, i.e., a larger lattice volume. The VMC and VPI 
results almost match for (000) and $(00\,-1)$ and are close to each other for 
(011) and (111). There is good agreement between the occupancies calculated by 
Eq.(\ref{eq:occupancy}) and those obtained by counting. This indicates that the 
single-particle wave function in Eq.(\ref{eq:phi_similar_to_Li}) is suitable 
to describe the bosons at each lattice site.

\subsection{Correlations}

\begin{figure}[t!]
\includegraphics*[width=8.5cm,viewport=167 520 543 783,clip]{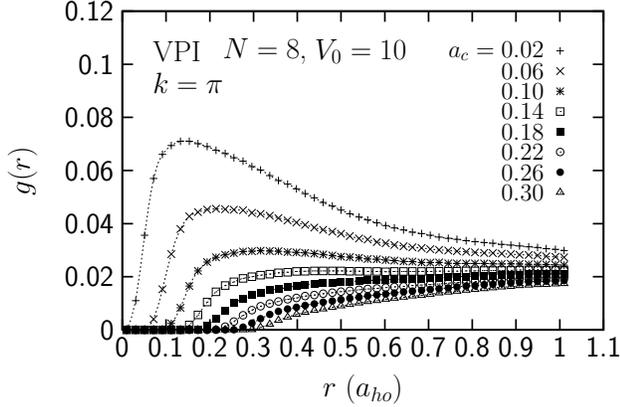}
\caption{VPI pair correlation function $g(r)$ for the indicated $a_c$. 
The system is the HS Bose gas of 
Fig.\ft\ref{fig:plot.ovrlp.vpi.N8.B10.severalk} at $k=\pi$.} 
\label{fig:gofr}
\end{figure}

\hs The VPI pair correlation functions $g(r)$ for the systems of 
$N=8$, $V_0=10$, $k=\pi$, and various $a_c$ in the range 
$0.02\le a_c \le 0.30$ are displayed in Fig.\ft\ref{fig:gofr}. 
It is observed that the correlations are strongest for $a_c=0.02$ 
and gradually weaken as $a_c$ is increased. 

\begin{figure}[t!]
\includegraphics*[width=8.5cm,viewport=175 455 540 710,clip]{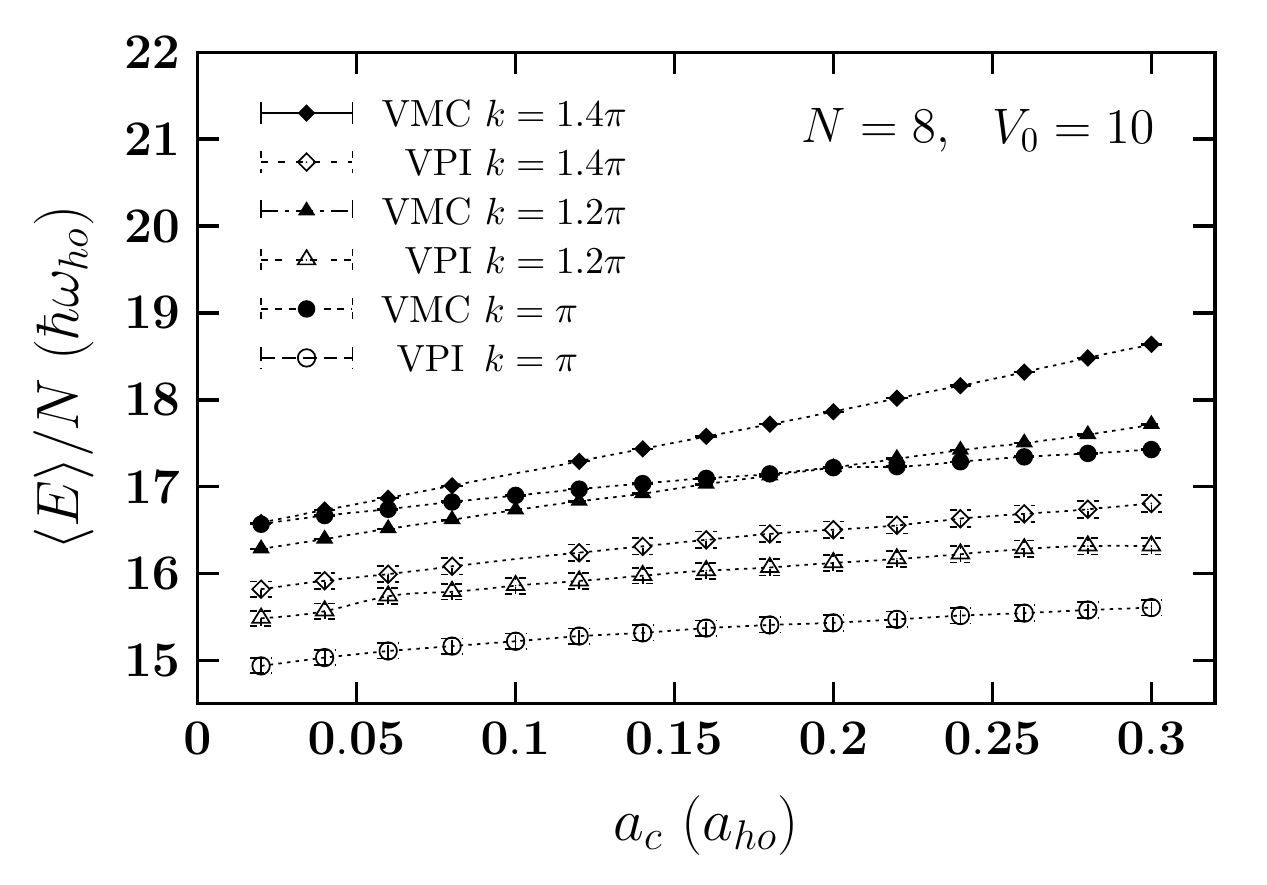}
\caption{Average energy per particle $\langle E\rangle/N$ vs. $a_c$ for
a HS Bose gas system of $N=8$, $V_0=10$, and the indicated values
of $k$. Solid and open diamonds: VMC and VPI results for $k=1.4\pi$.
Solid and open triangles: likewise, but for $k=1.2\pi$, and solid
and open circles: $k=\pi$.} 
\label{fig:plot.Energy.vs.ac.V10.N8.L27.severalk}
\end{figure}

\begin{figure}[t!]
\includegraphics*[width=8.5cm,viewport=173 456 535 710,clip]{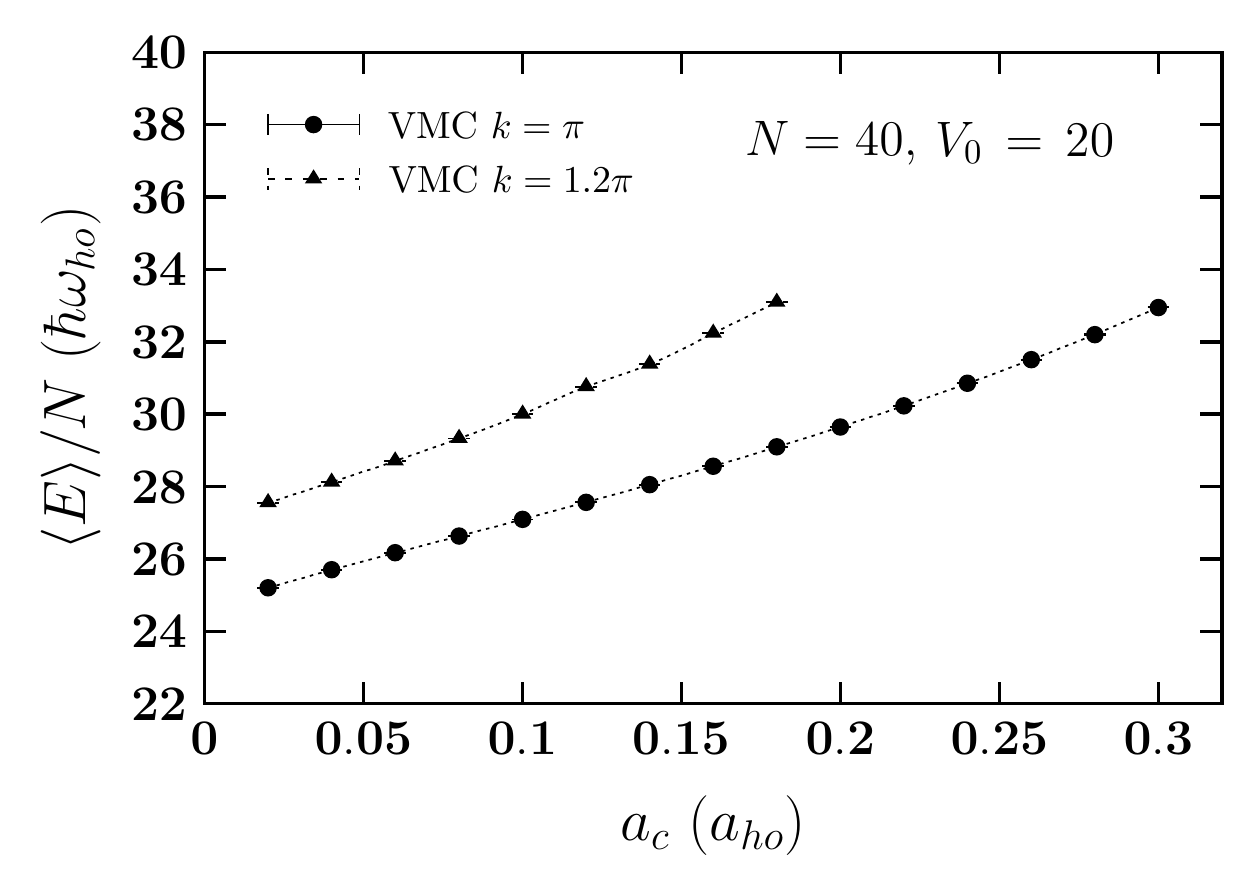}
\caption{As in Fig.\ft\ref{fig:plot.Energy.vs.ac.V10.N8.L27.severalk}; but for the HS Bose gases of 
Fig.\ft\ref{fig:plot.ovrlp.vmc.N40.B20.severalk}} 
\label{fig:plot.Energy.vs.ac.V20.N40.L27.severalk}
\end{figure}

\begin{figure}[t!]
\includegraphics*[width=8.5cm,viewport=173 458 535 711,clip]{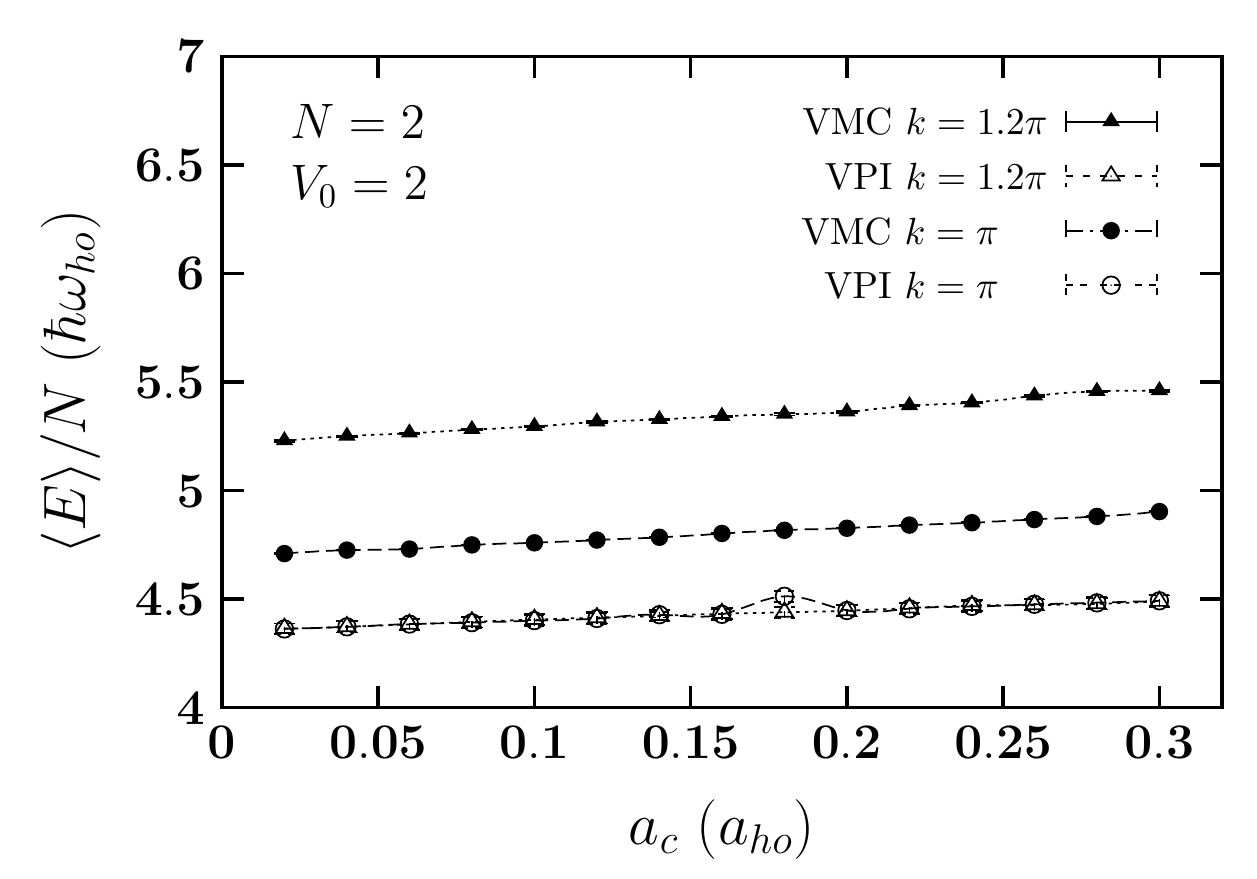}
\caption{As in Fig.\ft\ref{fig:plot.Energy.vs.ac.V10.N8.L27.severalk}; but
for $N=2$, $V_0=2$, and the indicated values of $k$.} 
\label{fig:plot.Energy.vs.ac.V2.N2.L27.severalk}
\end{figure}

\begin{figure}[t!]
\includegraphics*[width=8.5cm,viewport=162 216 444 695,clip]{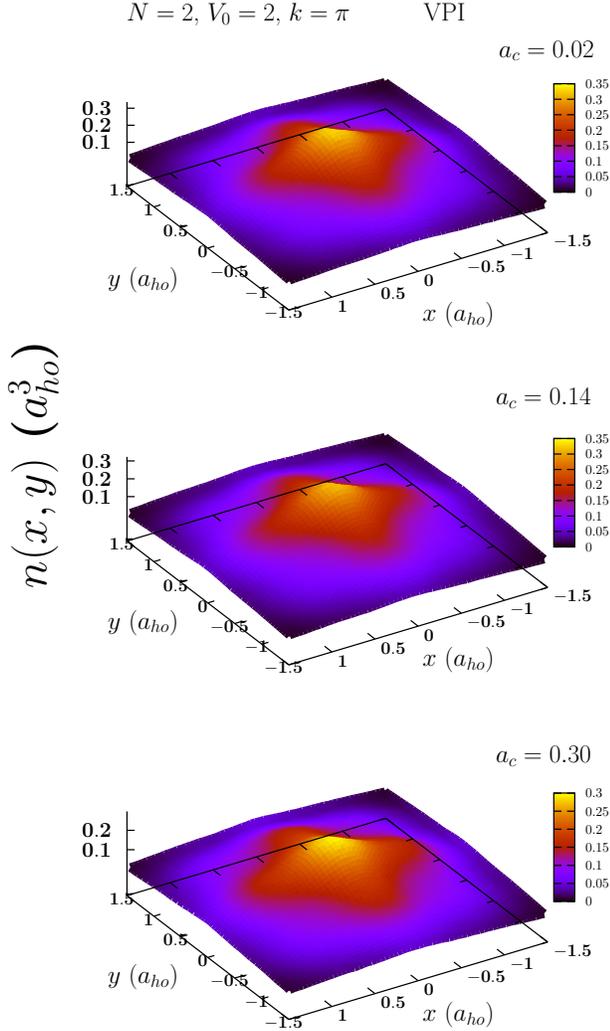}
\caption{Integrated VPI optical density for the HS Bose gas system of $N=2$, $V_0=2$, 
and $k=\pi$, at three values of $a_c$. From top to bottom frames: $a_c=0.02$, 0.14, 
and 0.30, respectively. Density and lengths are in trap units.} 
\label{fig:plot.od.vpi.V2.N2.L27.k1.0.figurestack}
\end{figure}

\begin{figure}[t!]
\includegraphics*[width=8.5cm,viewport=160 313 437 775,clip]{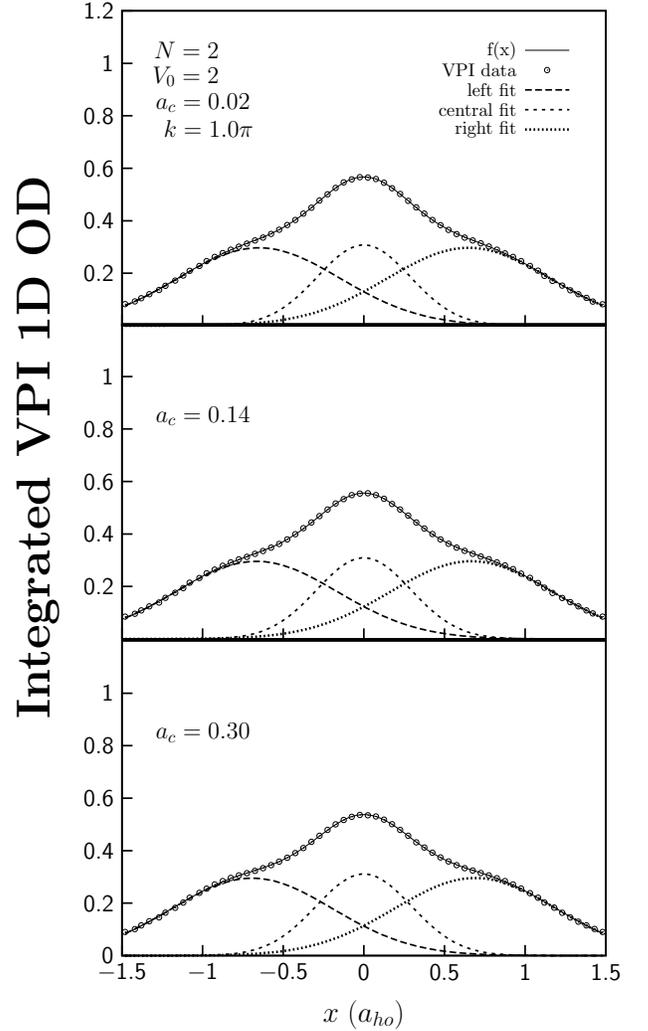}
\caption{Integrated one-dimensional optical density slices
of Fig.\ft\ref{fig:plot.od.vpi.V2.N2.L27.k1.0.figurestack} along
the x-axis. Solid line: fitting function $f(x)=F(x;X_1,X_2,X_3)$ 
Eq.(\ref{eq:total-fitting-function}), open circles: VPI data, 
thick dashed line: left fit $|\phi_{fit}(x,X_1)|^2$, thin dashed 
line: central fit $|\phi_{fit}(x,X_2=0)|^2$, dotted line: right 
fit $|\phi_{fit}(x,X_2)|^2$. From top to bottom, same $a_c$
at in Fig.\ft\ref{fig:plot.od.vpi.V2.N2.L27.k1.0.figurestack}.}
\label{fig:plot.ODx.vpi.vs.ac.B2N2k1.0ICTP.figurestack}
\end{figure}

\begin{figure}[t!]
\includegraphics*[width=8.5cm,viewport=153 205 443 689,clip]{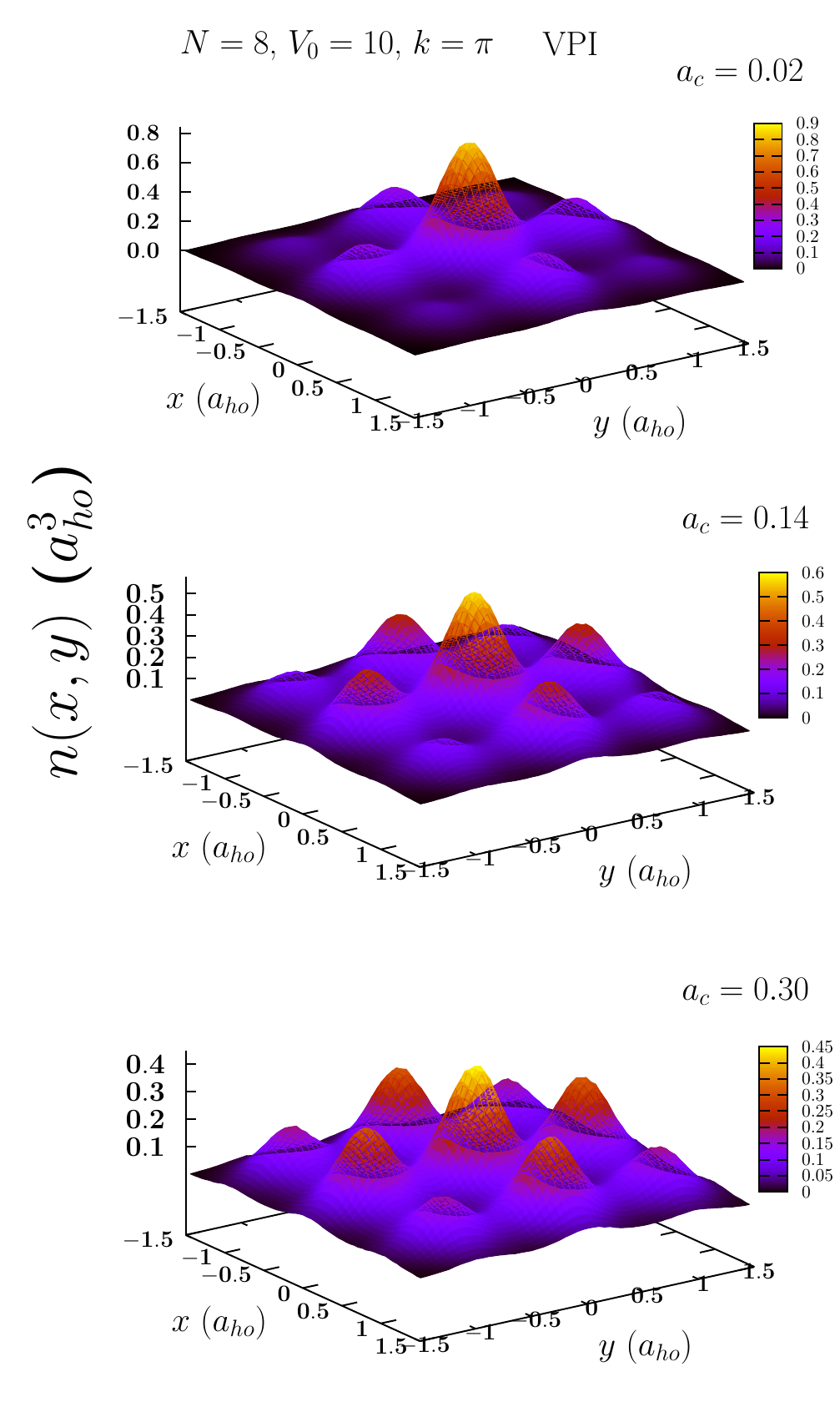}
\caption{As in Fig.\ft\ref{fig:plot.od.vpi.V2.N2.L27.k1.0.figurestack}; 
but for $N=8$ and $V_0=10$.} 
\label{fig:plot.od.vpi.V10.N8.L27.k1.0.figurestack}
\end{figure}

\begin{figure}[t!]
\includegraphics*[width=8.5cm,viewport=162 308 435 775,clip]{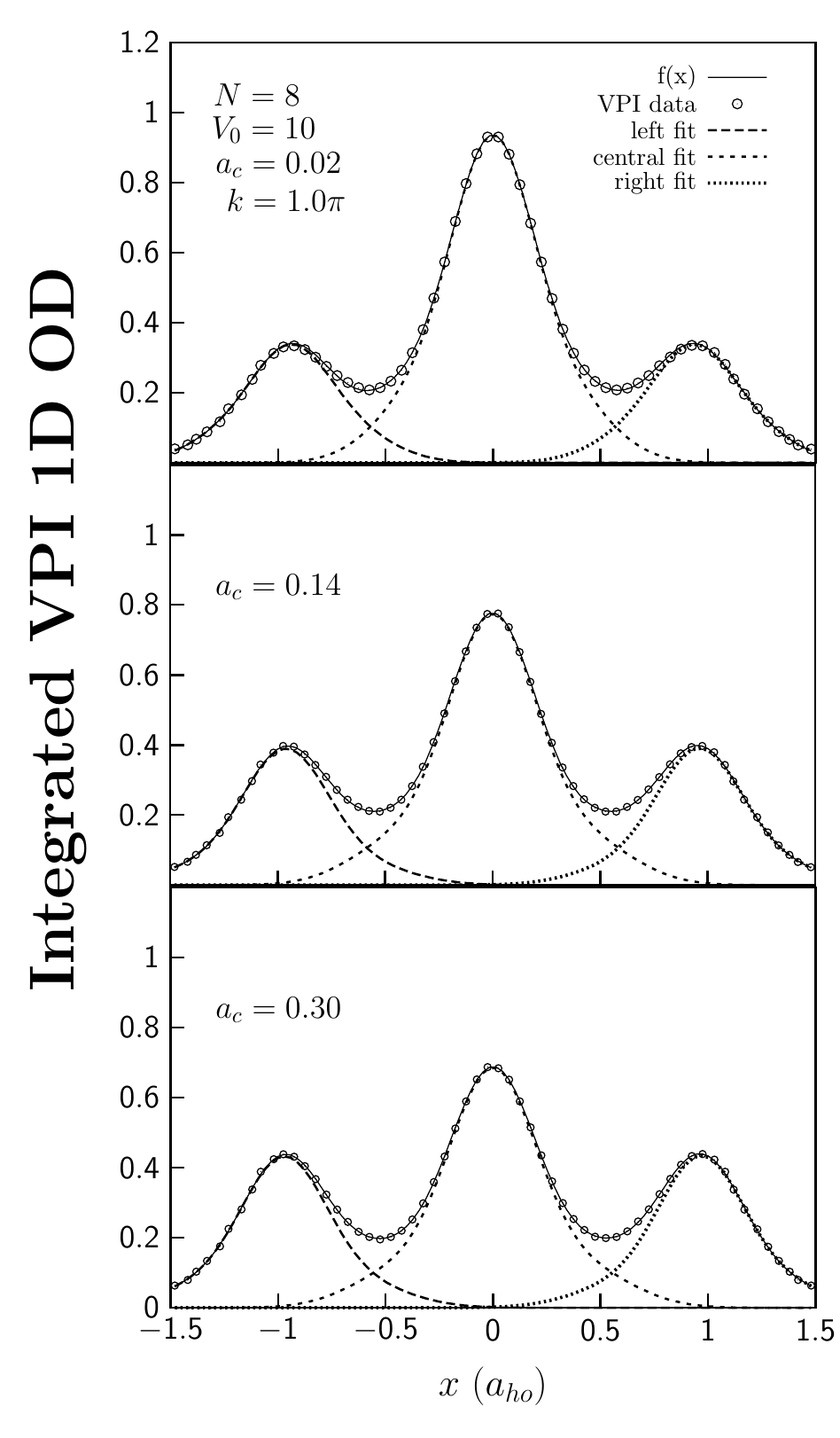}
\caption{As in Fig.\ft\ref{fig:plot.ODx.vpi.vs.ac.B2N2k1.0ICTP.figurestack}
but for $N=8$ and $V_0=10$.} 
\label{fig:plot.ODx.vpi.vs.ac.B10N8k1.0ICTP.figurestack}
\end{figure}

\begin{figure}[t!]
\includegraphics*[width=8.5cm,viewport=152 263 442 712,clip]{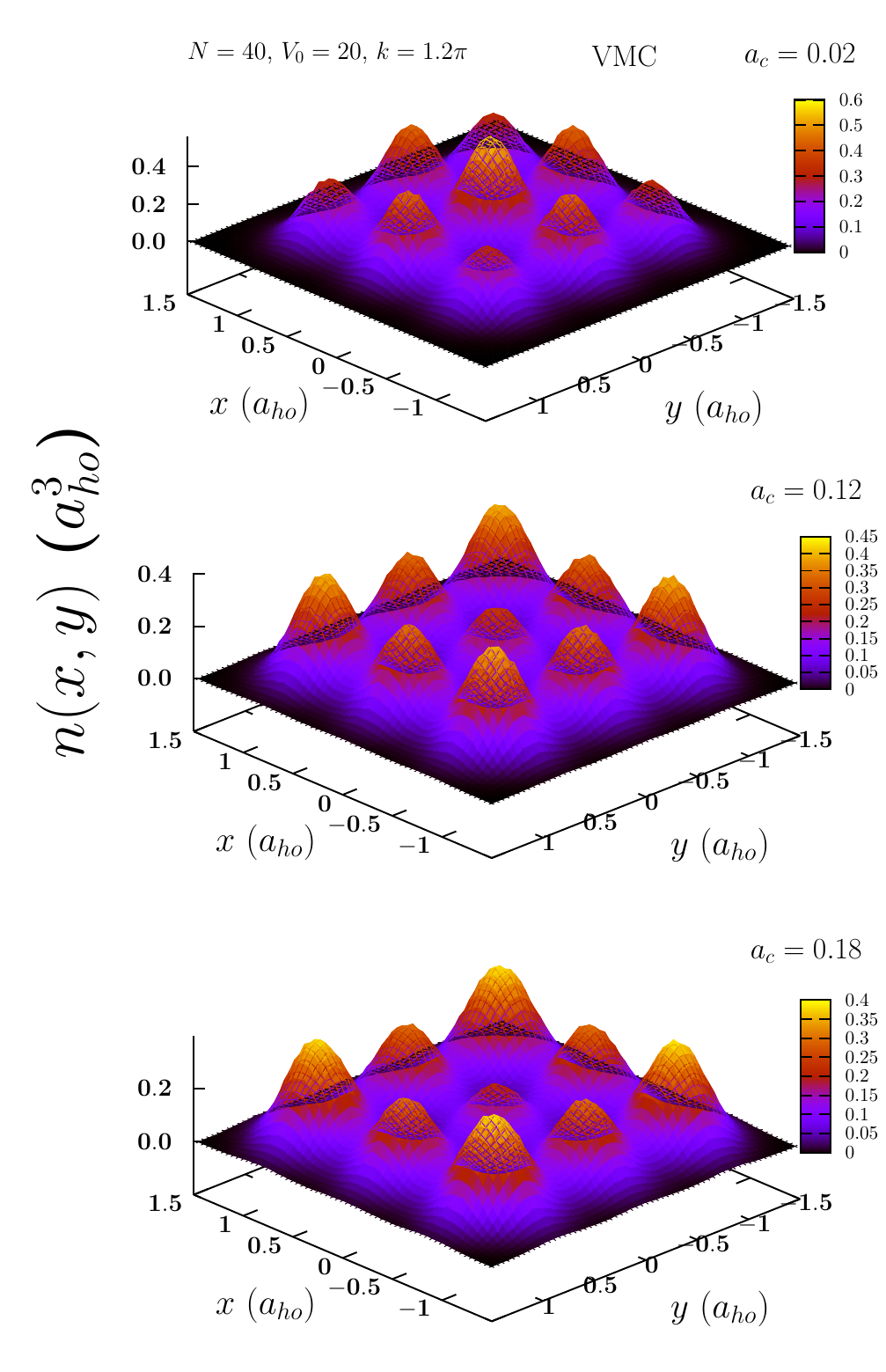}
\caption{As in Fig.\ft\ref{fig:plot.od.vpi.V2.N2.L27.k1.0.figurestack};
but for $N=40$, $V_0=20$, and $k=1.2\pi$.} 
\label{fig:plot.od.vmc.V20.N40.L27.k1.2.figurestack}
\end{figure}

\begin{figure}[t!]
\includegraphics*[width=8.5cm,viewport=160 311 433 773,clip]{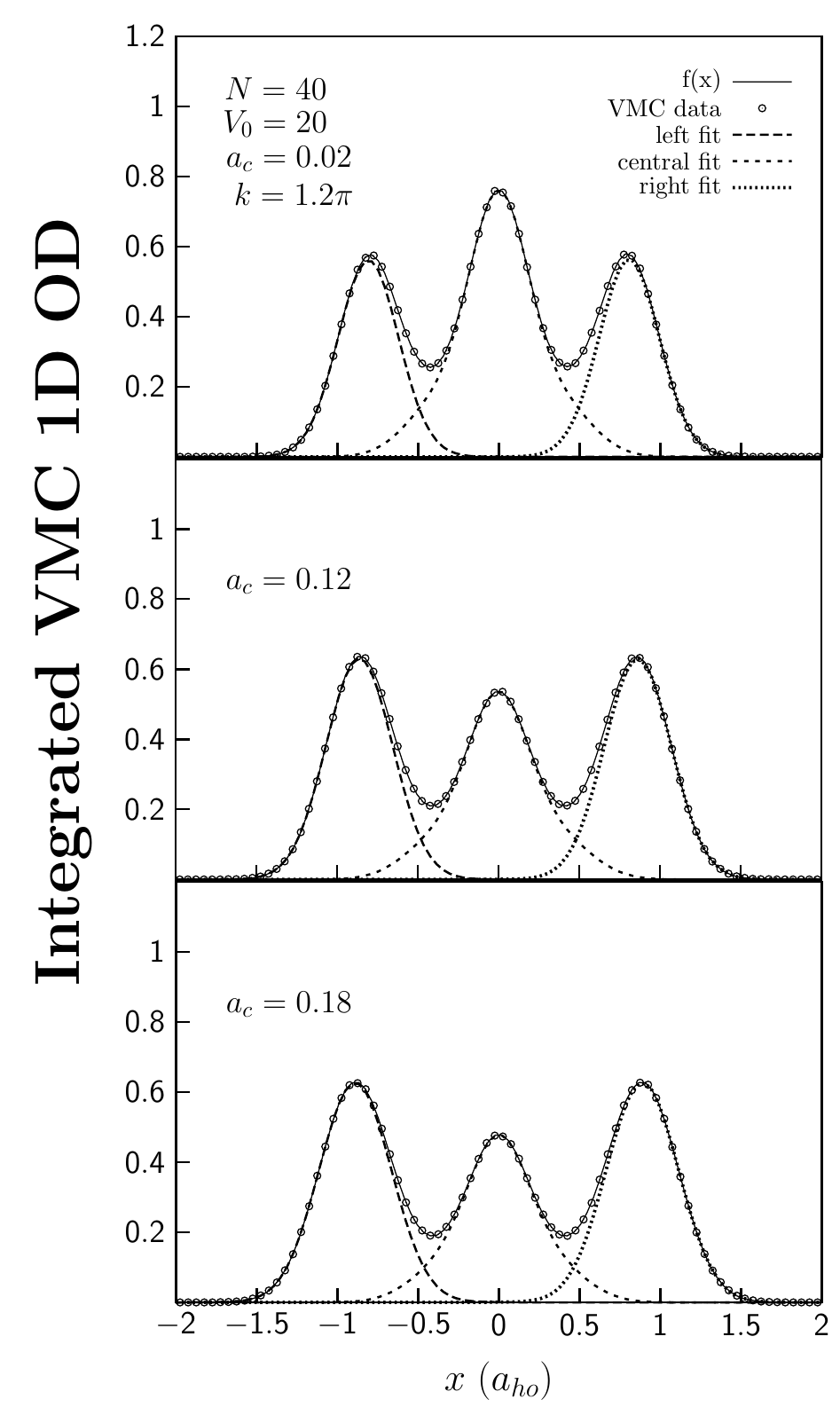}
\caption{As in 
Fig.\ft\ref{fig:plot.ODx.vpi.vs.ac.B2N2k1.0ICTP.figurestack} but
for $N=40$, $V_0=20$, and $k=1.2\pi$} 
\label{fig:plot.ODx.vmc.vs.ac.B20N40k1.2.figurestack}
\end{figure}

\begin{figure}[t!]
\includegraphics*[width=8.5cm,viewport=154 212 445 706,clip]{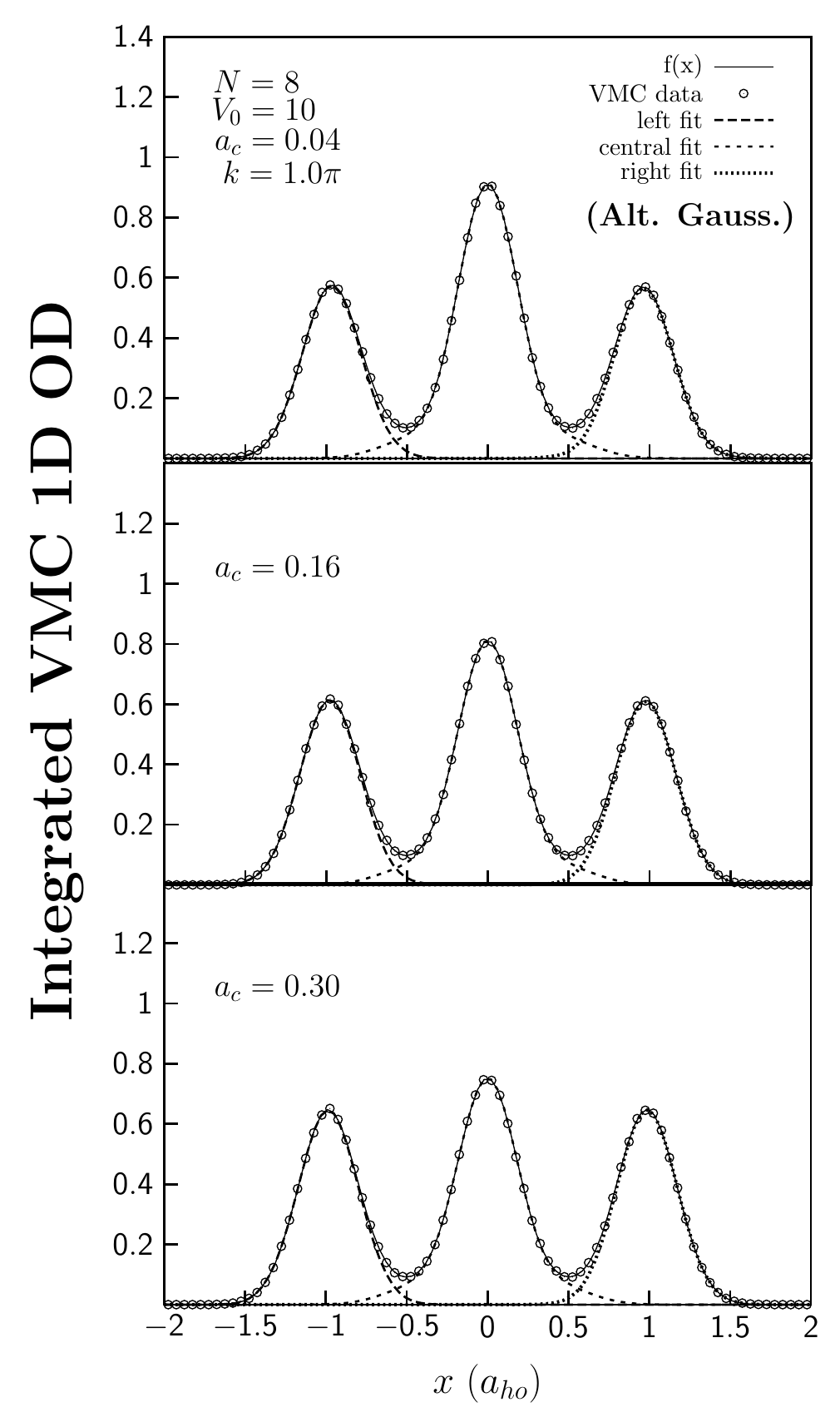}
\caption{As in Fig.\ft\ref{fig:plot.ODx.vpi.vs.ac.B10N8k1.0ICTP.figurestack};
but using the alternative Gaussian in Eq.(\ref{eq:alternative-Trial-wave-function}) 
({\bf Alt. Gauss}).} 
\label{fig:plot.ODx.vmc.vs.ac.V10.N8.L27.k1.0.altGauss.figurestack}
\end{figure}

\subsection{Energy}

\hs The average Monte Carlo (MC) energies per particle $\langle E\rangle/N$ as 
functions of $a_c$ for the systems of 
Figs.\ft\ref{fig:plot.ovrlp.vpi.N8.B10.severalk} 
and \ref{fig:plot.ovrlp.vmc.N40.B20.severalk}, are displayed in 
Figs.\ft\ref{fig:plot.Energy.vs.ac.V10.N8.L27.severalk}, and 
\ref{fig:plot.Energy.vs.ac.V20.N40.L27.severalk}, respectively. 
Additionally, the energies for a system of $N=2$ and $V_0=2$, are 
displayed in Fig.\ft\ref{fig:plot.Energy.vs.ac.V2.N2.L27.severalk} 
as well. The same legends are used in the latter three figures as follows. 
Solid and open diamonds: VMC and VPI results for $k=1.4\pi$, respectively. 
Similarly for triangles and circles: $k=1.2\pi$ and $\pi$, respectively. 
We note that, for most of the systems considered, $\langle E\rangle/N$ 
changes almost linearly within the given range of $a_c$ at a much slower 
rate than HS Bose gases in pure harmonic traps \cite{Dubois:01,Sakhel:02}. 
For example for $N=2$ with $k=\pi$, the VPI $\langle E\rangle/N$ rises only by 
$\Delta E_{VPI}/N\sim +2.98\%$ from $a_c=0.02$ to $a_c=0.3$, despite the fact 
that there is a very large change in $a_c$ (by a factor of 14!). Similarly for 
$N=8$ and $k=\pi$, $\Delta E_{VPI}/N$ is $\sim +4.48\%$. However, $\Delta E_{VPI}/N$ 
for $k=1.4\pi$ is more pronounced, being $\sim +6.26\%$. The corresponding VMC 
results are as follows: for $N=2$ with $k=\pi$, $\Delta E_{VMC}/N\sim +4.48\%$, 
for $N=8$ with $k=\pi$, $\Delta E_{VMC}/N\sim +5.20\%$, and for $N=8$ with 
$k=1.4\pi$, $\Delta E_{VMC}/N\sim +12.37\%$, respectively. For $N=40$, $\Delta E/N$ 
changes at a much faster rate than for a lower $N$. Further, the energy 
$\langle E_{VMC}\rangle/N$ rises with increasing $k$. 

\subsection{Optical density}\label{sec:optical-density}

\hs In this section, the integrated VPI optical density (OD) is 
displayed at different $a_c$. The goal is to explore the effects of 
variations in the interactions on the optical density profiles and on 
the tunneling amplitude. Particularly we show visually that, for a 
small number of particles, the width of the wave function in each 
well does not broaden with an increase in $a_c$ even up to order 0.1 and
that the overlap between the wave functions does not change either.

\hs Fig.\ft\ref{fig:plot.od.vpi.V2.N2.L27.k1.0.figurestack} displays 
the integrated VPI OD for $N=2$, $V_0=2$, and $k=\pi$. From the top to 
bottom frames, $a_c=0.02$, 0.14, and 0.30, respectively. As expected, 
the amplitude of the density is maximal at the center of the trap
and lowest for the lattice sites near the edges of the trap. Further,
the tunneling of the system into the potential barrier of the external
harmonic trap is suppressed due to the steep rise of this barrier
as one goes away from the center of the trap. Since the 
optical depth is very shallow, the wave function of the system is spread 
over the whole lattice, indicating the dominance of the tunneling process 
\cite{Greiner:02} over the localization effects of the optical lattice 
potential wells. Superflow in these systems is thus prevailent. 

\hs Fig.\ft\ref{fig:plot.ODx.vpi.vs.ac.B2N2k1.0ICTP.figurestack} displays 
one-dimensional density slices of 
Fig.\ft\ref{fig:plot.od.vpi.V2.N2.L27.k1.0.figurestack} along the x-axis, 
and additionally three functions of the form (\ref{eq:fitting-function}) 
fitted to this density profile. The fits reveal the extent of the overlap 
between the wave functions in neighboring wells. It is observed, that this 
overlap does not change with a rise in $a_c$. The left (dashed line) and
right (dotted line) density functions are much broader than the central 
function (thin dashed line) and all three peaks have almost the same 
amplitude. In fact, the overlap here is between three wave functions as 
those at the edges are able to spread over several trap lengths.
 
\hs Fig.\ft\ref{fig:plot.od.vpi.V10.N8.L27.k1.0.figurestack} is the same as 
Fig.\ft\ref{fig:plot.od.vpi.V2.N2.L27.k1.0.figurestack}; but for $N=8$ and 
$V_0=10$. Here, the bosons are more localized at the lattice sites, and the 
tunneling process is less dominant than the localization effects. Again, 
the effects of the external trap are manifested: the density diminishes 
towards the edges of the trap. 
Fig.\ft\ref{fig:plot.ODx.vpi.vs.ac.B10N8k1.0ICTP.figurestack} displays again the
one-dimensional slices of 
Fig.\ft\ref{fig:plot.od.vpi.V10.N8.L27.k1.0.figurestack} along the x-axis. Here, 
the overlap is much less pronounced than in 
Fig.\ft\ref{fig:plot.ODx.vpi.vs.ac.B2N2k1.0ICTP.figurestack}, and the 
amplitude of the central peak is larger than the peaks at the edges. The
left and right peaks do not overlap as in 
Fig.\ft\ref{fig:plot.ODx.vpi.vs.ac.B2N2k1.0ICTP.figurestack}. 
It seems that a slight increase in the number of particles
has a profound effect on the density profiles in that it changes the ratio
of the amplitude of the central density to that of the density at the trap
edges. Further, the width of the densities at the edges of the trap for $N=8$ 
has dropped by almost 2 trap lengths as compared to $N=2$. Nevertheless, 
there is still overlap between the wave functions in the neighboring wells 
even at high repulsion. For $N=8$, the amplitude of the central peak declines 
with increasing $a_c$. The change in the height of the central peak as one 
increases $a_c$ from 0.02 to 0.14 is $\sim 33.3\%$. Remarkably, even with 
strongly repulsive bosons, the density of the system in each well largely 
peaks at the potential minimum of each well instead of being broader and 
more evenly distributed around each lattice site as in 
Fig.\ft\ref{fig:plot.od.vpi.V2.N2.L27.k1.0.figurestack}.

\hs The latter situation does not change very much for a larger number
of particles. Fig.\ft\ref{fig:plot.od.vmc.V20.N40.L27.k1.2.figurestack} 
displays the integrated optical density for the system of $N=40$,
$V_0=20$, and $k=1.2\pi$ and 
Fig.\ft\ref{fig:plot.ODx.vmc.vs.ac.B20N40k1.2.figurestack} its density
slices, as in the latter figures for the optical density. Again, the
peaks in each well do not broaden very much with a rise in $a_c$.
The central peak looses amplitude in favor of the peaks at the edges.
On closely inspecting the area of the overlap between neighboring
peaks, one can depict a small decline in the tunneling amplitude
of the system as $a_c$ is increased. In 
Figs.\ft\ref{fig:plot.ODx.vpi.vs.ac.B10N8k1.0ICTP.figurestack} and 
\ref{fig:plot.ODx.vmc.vs.ac.B20N40k1.2.figurestack}, the positions 
of the left and right peak maxima shift to the left and right, 
respectively, as $a_c$ is increased.

\subsection{Alternative VMC Trial Function}

\hs In order to test the validity of using a Gaussian ($\exp(-\alpha r^2)$) 
in our trial wave function Eq.(\ref{eq:Trial-wave-function}), we calculated 
the optical densities for $N=8$, $V_0=10$, and $k=\pi$ using a slightly 
modified version of our trial wave function:

\begin{eqnarray}
&&\Psi(\{\mathbf{r}\}, \{\mathbf{R}\})\,=\,\nonumber\\
&&\prod_{i=1}^N\,\exp(-\alpha\,r_i^2-\epsilon\,r_i^4)\,
\psi(\mathbf{r}_i,\{\mathbf{R}\})\,\prod_{i < j} f(|\mathbf{r}_i-\mathbf{r}_j|), \nonumber\\
\label{eq:alternative-Trial-wave-function}
\end{eqnarray}

with an additional parameter $\epsilon$ added to the previous set of variational
parameters. Nevertheless, this produces similar results to the previous
section. 
Fig.\ft\ref{fig:plot.ODx.vmc.vs.ac.V10.N8.L27.k1.0.altGauss.figurestack}
displays the density slices for the system of $N=8$, $V_0=10$, and $k=\pi$, 
using Eq.(\ref{eq:alternative-Trial-wave-function}). The wave functions in
each well do not broaden, but the central peak looses amplitude with
increasing $a_c$, whereas the peaks at the edges gain amplitude.

\section{SF to MI pseudotransition}

\hs Although our chief goal was to fix $V_0$ and vary $a_c$ only, we
nevertheless divert a little and explore the evolution of a HS Bose 
gas in a CHOCL by changing $V_0$ and fixing $a_c$. The goal is to 
check whether there is a critical $V_0$ at which a SF to MI transition
occurs in the presence of an external harmonic trap.

\begin{figure}[t!]
\includegraphics*[width=8.5cm,viewport=168 163 433 775,clip]{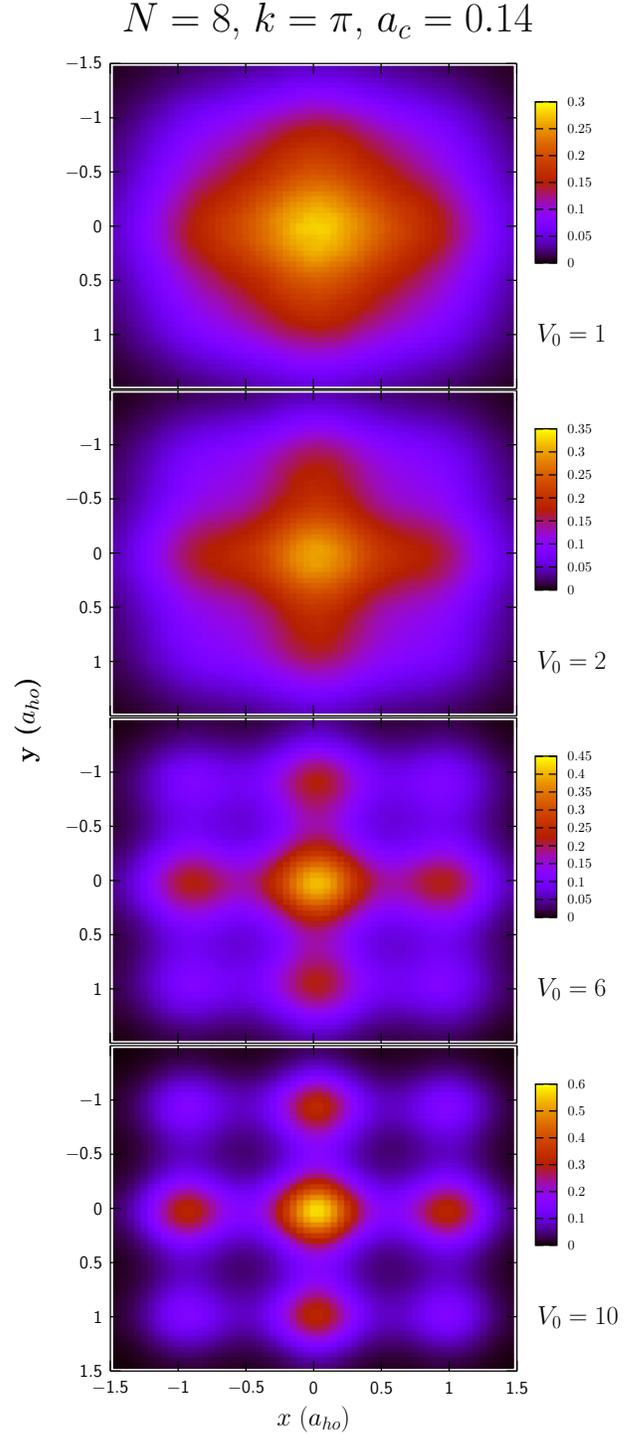}
\caption{Map view of integrated VPI optical density for
various optical depths $V_0$. The system is a HS Bose gas
of $N=8$, fixed $a_c=0.14$, and $k=1.0\pi$ which is confined
in a CHOCL. From top to bottom: $V_0=1$, 2, 6, and 10. Density
and lengths are in trap units.}
\label{fig:plot.od.vpi.N8.k1.0.severalV.pseudotransition}
\end{figure}

\begin{figure}[t!]
\includegraphics*[width=8.0cm,viewport=168 206 430 770,clip]{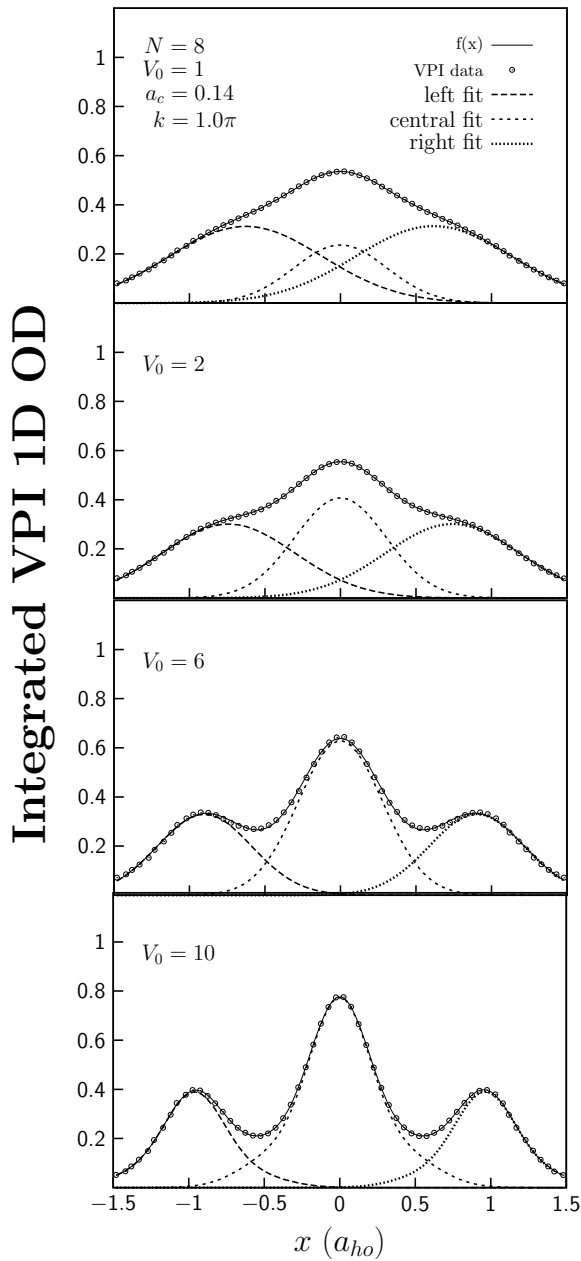}
\caption{Density slices of 
Fig.\ft\ref{fig:plot.od.vpi.N8.k1.0.severalV.pseudotransition}
along the x-axis. The legends describe the same functions
as in Fig.\ft\ref{fig:plot.ODx.vpi.vs.ac.B2N2k1.0ICTP.figurestack}}.
\label{fig:plot.ODx.vpi.vs.V0.N8ac0.14k1.0ICTP.figurestack}
\end{figure}

\hs Fig.\ft\ref{fig:plot.od.vpi.N8.k1.0.severalV.pseudotransition} 
displays the evolution of the integrated optical density (map view)
for a system of $N=8$ particles, $k=\pi$, fixed HS diameter 
$a_c=0.14$, and various optical depths $V_0$. From top to bottom:
$V_0=1$, 2, 6, and 10, respectively. As $V_0$ is increased, the 
overlap of the wave functions in neighboring wells drops, however,
the change is not abrupt but {\it gradual}. That is, there does
not exist a critical point for a superfluid to Mott-insulator
transition. Fig.\ft\ref{fig:plot.ODx.vpi.vs.V0.N8ac0.14k1.0ICTP.figurestack}
displays, similarly to 
Figs.\ft\ref{fig:plot.ODx.vpi.vs.ac.B2N2k1.0ICTP.figurestack} and 
\ref{fig:plot.ODx.vpi.vs.ac.B10N8k1.0ICTP.figurestack}, 
again the density slices of 
Fig.\ft\ref{fig:plot.od.vpi.N8.k1.0.severalV.pseudotransition} along
the x-axis, where one can clearly see how the overlap, and therefore 
the tunneling, drops. We anticipate then, that as $V_0$ is increased 
further, the tunneling will drop substantially. Note, however, that 
the amplitude of the central peak rises with increasing $V_0$.

\section{Discussion and conclusion}\label{sec:discussion}

\hs In summary, we have investigated the effects of strongly repulsive 
interactions on the tunneling amplitude of HS bosons confined in a 
3D${}^{\hbox{\it al}}$ simple cubic optical lattice plus by a tight 
spherical harmonic trap. The tunneling amplitude was measured by the overlap
integral of the wave functions in neighboring wells. The external harmonic 
trap was introduced in order to cause an inhomogeneiety in the density 
distribution of the trapped system. Another goal was also to explore the 
effect of this inhomogeneiety on the SF to MI transition of bosons in 
optical lattices. 

\hs It was found that for small N, the tunneling amplitude does not change 
by increasing the boson-boson repulsion at fixed lattice spacing $d=\pi/k$. 
This is mainly because the width of each localized wave function does not 
broaden with a rise in the repulsive forces. The tunneling amplitude, however, 
rises by reduction of $d$ at fixed $a_c$. Even at very large repulsion, 
tunneling is still present in the system and does not vanish completely. 
The presence of an external trap seems to suppress the SF to MI abrupt transition. 
The transition is gradual and there is no criticality. 

\hs The energies change only weakly with increasing $a_c$ for $N=2$, but a 
more pronounced change is observed  for $N=8$, and is much larger for $N=40$. 
The occupancy of the central and first nearest neighbor sites drops with 
increasing $a_c$, and rises at the second and third nearest neighbor sites.

\subsection{Occupancy and onsite interaction energies}

\hs Although the occupancy of particles in 
Fig.\ft\ref{fig:plot.occupN.vs.ac.V10.N8.L27.severalk.figurestack} at 
sites $(000)$ and $(00\,-1)$ drops with increasing $a_c$ causing a drop in 
the density at the central lattice site, the corresponding 
$\langle U_{(000)}\rangle$ and $\langle U_{(00-1)} \rangle$ still rise. Thus the 
effect of increasing $g$ via $a_c$ on $\langle U_n\rangle$ overwhelms that of the 
drop in the occupancy at each lattice site. The net result is that 
$\langle U_{(pqr)}\rangle$ increases with increasing $g$. Further, the occupancy 
of the lattice sites plays a crucial role in determining the relative magnitudes 
of $\langle U_{(pqr)}\rangle$. Since (000) has the highest occupancy at all $a_c$, 
in Figs.\ft\ref{fig:plot.U.for.4.sites.vs.ac.V10.N8.k1.0.L27} and 
\ft\ref{fig:plot.U.for.4.sites.vs.ac.V10.N8.k1.2.L27} it has also the highest 
onsite repulsion. Then with decreasing occupancy, the onsite repulsion decreases 
in the order $U_{(000)}$, $U_{(00\,-1)}$, $U_{(011)}$, and $U_{(111)}$ at all $a_c$.
In Fig.\ft\ref{fig:plot.occupN.vs.ac.V10.N8.L27.severalk.figurestack}, one can 
also see that (000) has the highest, whereas (111) the lowest occupancy.

\hs In equilibrium, one could assume that there are as many particles 
tunneling into a lattice site as out which preserves the occupancy at each 
lattice site in equilibrium \cite{Mullin:09}. When $a_c$ is changed, the 
particles relocate themselves in order to reach a new equilibrium configuration. 
If $a_c$ is increased, particles are pushed out of the central site and occupy 
others until a new equilibrium configuration is reached again. Due to the 
external trap they are not able to tunnel further away from the edges of the 
trap. Further, note it is possible that during the equilibration process the 
particles having tunneled from the center of the trap to the first neighbors 
continue tunneling to the second and third. In fact, it would be interesting 
to know how the particles' motion is channelled along the sites of a larger 
simple cubic optical lattice, as the repulsion between the bosons is increased. 
Finally, the average (noninteger) occupation numbers $\langle N_{(pqr)}\rangle$ 
at each lattice signal that the systems are SF. This indicates also that the 
particles are fragmented all throughout the lattice.

\subsection{Optical density and tunneling}\label{sec:SF-to-MI-pseudo-transition}

\hs As we have seen in 
Figs.\ft\ref{fig:plot.od.vpi.V2.N2.L27.k1.0.figurestack} -
\ref{fig:plot.ODx.vmc.vs.ac.B20N40k1.2.figurestack}, as much as
the repulsion was increased, the overlap between neighboring
wells did not change significantly, even at large $N$. But the
change in the overlap is more pronounced if one increases $V_0$ 
while keeping $a_c$ fixed. This was the situation displayed in 
Fig.\ft\ref{fig:plot.ODx.vpi.vs.V0.N8ac0.14k1.0ICTP.figurestack}.
We conclude then, that a MI state can be obtained more efficiently by 
rather increasing the optical depth $V_0$ than the HS repulsion 
between the bosons, $a_c$.

\hs Similarly to our findings in Sec.\ref{sec:optical-density}, Greiner \ea\ 
\cite{Greiner:02} showed, by applying the Bose-Hubbard (BH) model to a 3D 
optical lattice, that if the tunneling dominates the BH Hamiltonian, the 
single particle wave functions are spread over the whole lattice, and phase 
coherence existing between the lattice sites forms a SF. On the other hand, if 
the atom-atom interactions dominate, there is no phase coherence. Therefore, 
the single-particle wavefunctions become localized at their lattice sites with 
a fixed number of atoms, thus forming a MI state. In our case, however, we do not 
get a {\it pure} MI even at very high repulsion because tunneling is still 
present there. In the highly repulsive regime, we may nevertheless be talking
about a {\it quasi} MI state where tunneling is present but vanishingly small.

\hs In Fig.\ft\ref{fig:plot.od.vpi.N8.k1.0.severalV.pseudotransition}, 
if one keeps increasing $V_0$, we anticipate that the tunneling amplitude 
will eventually drop to zero signaling initially what looks like an entrance 
into the MI regime. However, in order to identify a pure MI regime, integer 
occupancy of each lattice site is needed. Because the number of particles is 
less than the number of lattices sites ($N/N_L < 1$), the particles are 
fragmented \cite{Konotop:02} and we do not get integer occupancy at each lattice site. 
This was shown in Sec.\ref{sec:occupancy-results} and as a result this 
system may be regarded a quasi MI with a small SF component.

\subsection{The role of the lattice spacing $d=\pi/k$}\label{sec:lattice spacing}

\hs In Figs.\ft\ref{fig:plot.ovrlp.vpi.N8.B10.severalk}, and 
\ref{fig:plot.ovrlp.vmc.N40.B20.severalk}
the reduction in $\pi/k$ reduces the width of the optical potential barriers, 
thus reducing the distance a particle needs in order to tunnel completely 
through a barrier. As a result, an increase in the overlap occurs between 
two neighboring single-particle wavefunctions $\phi(\mathbf{r},\mathbf{R}_n)$ 
and $\phi(\mathbf{r},\mathbf{R}_{n+1})$ causing a rise in \Ip. The increase in \Ip\ 
with $k$ could be further explained on the basis of a recent evaluation of the tunneling 
rate in 1D through a single, tilted optical barrier without external confinement by 
Huhtam\"aki \ea\ \cite{Huhtamaeki:07}. If $a$ and $b$ are the left-hand and right-hand 
classical turning points of a potential barrier, then if the distance between $a$ and 
$b$ is reduced by increasing the lattice wave vector $k$, the overlap between the wave 
functions in neighboring wells rises.

\subsection{The effects of the external harmonic trap}\label{sec:harmonic-trap}

\hs The mobility of the bosons is eventually restricted to the confining 
volume of the external harmonic trap. By increasing the size of the bosons 
with $a_c$, the available free space for motion in the confined volume is reduced. 
Further, as the repulsion rises with $a_c$, the tunneling amplitude remains
constant for a small number of particles as they get locked inside the lattice
sites.
\hs A tight external harmonic potential also suppresses the critical SF to
{\it pure} MI transition as was shown in 
Fig.\ft\ref{fig:plot.ODx.vpi.vs.V0.N8ac0.14k1.0ICTP.figurestack}. 
In fact, Gygi \ea\ \cite{Gygi:06} argued similarly that the presence of a 
gradient due to the external confining potential destroys quantum criticality 
in the transition from the SF to the MI phase; this gradient is present in our 
simulations. Further, as high as the repulsion between the bosons in our systems 
becomes, tunneling is still present due to the presence of external confinement. 
This forces an inhomogeneiety in particle distribution over the lattice sites. 

\subsection{Energy}

\hs The rise in $\langle E/N\rangle$ with $k$ in 
Figs.\ref{fig:plot.Energy.vs.ac.V10.N8.L27.severalk},  
\ref{fig:plot.Energy.vs.ac.V20.N40.L27.severalk}, and
\ft\ref{fig:plot.Energy.vs.ac.V2.N2.L27.severalk},
could be attributed to Heisenberg's uncertainty principle.
A decline in the available volume the bosons can occupy in the optical lattice 
wells reduces the uncertainty in their positions. Thus, the uncertainty in their 
momenta increases and as a result their kinetic energy (quantum pressure) rises. 
It can also be seen that a change in $k$ has a more profound effect on $E$ than 
a change in $a_c$. Therefore, the lattice dimension seems to play a profound role 
in determining the SF properties of bosons in optical lattices in that it
can substantially control the tunneling amplitude between neighboring wells.

\subsection{Other work}

\hs Similarly to Bach and Rz$\dot{a}$zewski \cite{Bach:04}, we used 
Gaussians of the form $\exp[-\alpha(x-x_n)^2]$ but weighted by a polynomial 
as in Eq.\ft(\ref{eq:phi_similar_to_Li}). These authors determined that 
the mobility of the atoms, as described by the hopping parameter $t$ and 
their interactions $U$, is controlled by the optical-lattice depth. If this 
depth is shallow, the atoms become delocalized over the whole lattice as 
it almost occurs in Fig.\ft\ref{fig:plot.od.vpi.V2.N2.L27.k1.0.figurestack}. 

\hs Li \ea\ \cite{Li:06} investigated the SF to MI transitions in atomic 
BECs confined in optical lattices by using the BH model. Using 
an isotropic cubic lattice, they have chosen a variational trial function 
of the form $g(u)\,=\,(1+\alpha u^2)\,\exp(-\beta u^2)$ (a Wannier function); 
our single-particle wave function is the same as theirs, except for an 
additional factor $\propto u^4$ [Eq.(\ref{eq:phi_similar_to_Li})]. They 
addressed the possibility of observing SF to MI transitions for an average 
lattice-site occupancy larger than one. It was noted that by increasing the 
repulsion between the lattice bosons, their wavefunction in each well broadens, 
thus enhancing $J$ between neighbouring lattice sites. In contrast, our atomic 
clouds did not expand due to the presence of external confinement. Li \ea\ 
evaluated the on-site interaction energy by variationally minimizing the energy 
with respect to the parameters $\alpha$ and $\beta$. 

\hs Capello \ea\ \cite{Capello:01} used VMC to describe the MI transition 
(MIT) by means of a variational Gutzwiller wavefunction in discrete space 
and optimized the variational parameters to achieve the ground state system. 
In contrast we used a wave function in continuous space and the total energy 
was minimized in our research as they did. 

\begin{center}{\bf ACKNOWLEDGEMENT }\end{center}

\hs Interesting and enlightening discussions with Saverio Moroni and William 
Mullin are gratefully acknowledged. We thank William Mullin for a critical reading 
of the manuscript. One of the authors (ARS), gratefully acknowledges the International 
Center for Theoretical Physics (ICTP) for providing computational access to
their GRID cluster facility.

\bibliography{./optical-lattice,./MonteCarlo}
\end{document}